\documentclass[10pt]{article}
\usepackage{graphicx}
\usepackage{setspace}
\usepackage{amssymb,amsmath}
\usepackage{natbib}
\usepackage{graphics}
\usepackage[normalem]{ulem}

\DeclareGraphicsExtensions{.pdf}

\begin{document}

\title{\hspace*{1cm}\vspace*{-4cm}\\Quantum scissors -- finite-dimensional states engineering }

\author{\small W. Leo\'nski$^\dagger$ and A. Kowalewska-Kud{\l}aszyk$^\ddagger$\vspace{0.5cm}\\
 $^\dagger$ \textit{\small Quantum Optics and Engineering Division, Institute of Physics,}\\
\textit{\small University of Zielona G\'ora, ul. Prof. Z.~Szafrana 4a, 65-516 Zielona G\'ora, Poland}\vspace*{0.5cm}\\
\textit{\small $^\ddagger$ Nonlinear Optics Division, Physics Department, A. Mickiewicz University}\\
\textit{\small ul. Umultowska 85.
61-614 Pozna\'n, Poland}}
\date{}
\maketitle
\begin{abstract}
This work discusses quantum states defined in a finite-dimensional Hilbert space. In particular, after the presentation of some of them and their basic properties the work concentrates on the group of the quantum optical models that can be referred  to as \textit{quantum optical scissors}. Such ''devices'' can generate on their outputs states that are finite-dimensional, and simultaneously use for such preparation quantum states that are defined in the infinity-dimensional space. The work concentrates on  two groups of models: the first one, comprising linear elements and the second one -- models for which optical, Kerr-like nonlinear elements were applied.
\end{abstract}

\section{Introduction}
Problems of quantum optical states engineering have attracted remarkable interest in last years. Various concepts of such states and methods of their production and manipulation have been presented in numerous papers. They have diverse applications in atomic and molecular, solid state and nano-systems physics, and also in the quantum information theory. The latter have recently given a stimulating pulse for the investigation of the states defined in finite-dimensional Hilbert space. However, one should keep in mind that the general idea of such states was born much earlier. In particular, Radclife \citep*{R71} and Arecchi \textit{et.al} \citep*{ACG72} proposed the atomic (or spin) coherent states definition for the optical models involving atomic systems interacting with transverse electromagnetic field. Those states are finite-dimensional analogues of the coherent states proposed by  Glauber \citep*{G63a,G63c} and play a crucial role in the quantum optics theory. Finite-dimensional coherent states have been discussed in various aspects -- for example see \citep*{BWK92,KWZ93,KWZ94,MPT94,MLI01,LM01} \textit{and the references quoted therein}.

Another milestone in the development of the idea of finite-dimensional states was the proposal of the \textit{phase-states} given by Pegg and Barnett in \citep*{PB88,PB89}. The key idea of the definition they proposed is to calculate the state and all physical quantities in the $(s+1)$-dimensional space and then, to take the limit $s\rightarrow\infty$. These states are not the subject of this paper and we shall concentrate on the quantum scissors systems, however, they are interesting enough to be mentioned here. For instance, in \citep*{VAS93} the model involving atoms injected into a cavity was discussed as a potential source of the \textit{phase-states}. The Pegg-Burnett formalism is important as a method of defining other finite-dimensional states.
As it will be presented, some of the finite-dimensional states are defined in an analogous way to the \textit{phase-states}, \textit{i.e.} it shall be assumed that the space is finite-dimensional and within such a space all operators and desired states are defined. Such states we will referred to as \textit{finite dimensional states}. Another approach to the states definiton which is presented in this paper, is similar to that proposed in \citep*{KWZ93,KWZ94}. For this case the expansion of the discussed state (for instance, coherent or squeezed ones) in $n$-photon basis derived in infinite-dimensional space and then truncation of this space applied. Obviously, the proper normalization should be performed in this procedure. To distinguish such defined states from the \textit{finite dimensional} ones they shall be called \textit{truncated states}.

Nowadays finite-dimensional states (FDS) seems to be especially important from the point of view of the quantum information theory. In particular, two- or multi-modes states are relevant and they are commonly discussed in a context of quantum entanglement. Examples of such states are Bell (\citep*{NH00,H06,LP07,D07,J07} \textit{and the references quoted therein}), GHZ (Greenberg-Horne-Zeilinger) \citep*{GHZ89} and $W$-states \citep*{DVC00} states. It should be emphasized, the problem of quantum entanglement is also one of the most intriguing problems of the current quantum theory and it is believed that its solution will give us a new insight into the nature of the quantum world. Therefore, the finite-dimensional and truncated states as those closely related to the entanglement problems, play a principal role in answering the fundamental questions of the quantum theory.

The paper is organized as follows. First part of this article is devoted to some introduction to the quantum optical states which are defined in finite-dimensional Hilbert space. Next, various methods of generation  such states are presented. The  presentation starts with the models involving linear optics elements only -- \textit{linear quantum scissors} (LQS). After that the work concentrates on the methods of using nonlinear optics elements. This family of the quantum optical models we refer  to as \textit{nonlinear quantum scissors} (NQS). In this paper mostly quantum optical models shall be discused, although some of them, due to their universality, can be applied in non-optical systems.

\section{Finite dimensional quantum states}

\subsection{$n$-photon states}
The presentation of FDS shall start from the simplest case of the $n$-photon \textit{Fock} states. These states are commonly used for the description of the quantum electromagnetic field and are widely discussed in the literature (for the properties of these states see for instance \citep*{P84,GK05,Twww} \textit{and the references quoted therein}).  
They are described as eigen-states of number of photons operator $\hat{n}$ defined by the boson creation and annihilation operators $\hat{a}^\dagger$ and $\hat{a}$, respectively
\begin{equation}
\hat{n}\,=\,\hat{a}^\dagger\hat{a}\quad 
\end{equation}
which can be expressed by the formula
\begin{equation}
\hat{n}|n\rangle\,=n|n\rangle\quad .
\end{equation}
In fact, mathematically  they are the same states as those describing quantum harmonic oscillator.

The  bosonic creation and annihilation operators act on the $n$-photon $|n\rangle$ states as follows
\begin{equation}
\begin{split}
\hat{a}|n\rangle\,&=\,\sqrt{n}|n-1\rangle\\
\hat{a}^\dagger |n\rangle\,&=\,\sqrt{n+1}|n+1\rangle\quad .
\end{split}
\end{equation} 
In consequence, the $n$-photon $|n\rangle$ state can be obtained by a successive action of the creation operator $\hat{a}^\dagger$ on the vacuum state $|0\rangle$. This action can be described by the formula
\begin{equation}
|n\rangle\,=\,\frac{(\hat{a}^\dagger)^n}{\sqrt{n!}}\,|0\rangle\quad .
\end{equation}

\begin{figure}[h!]
\begin{center}
\includegraphics[scale=0.35]{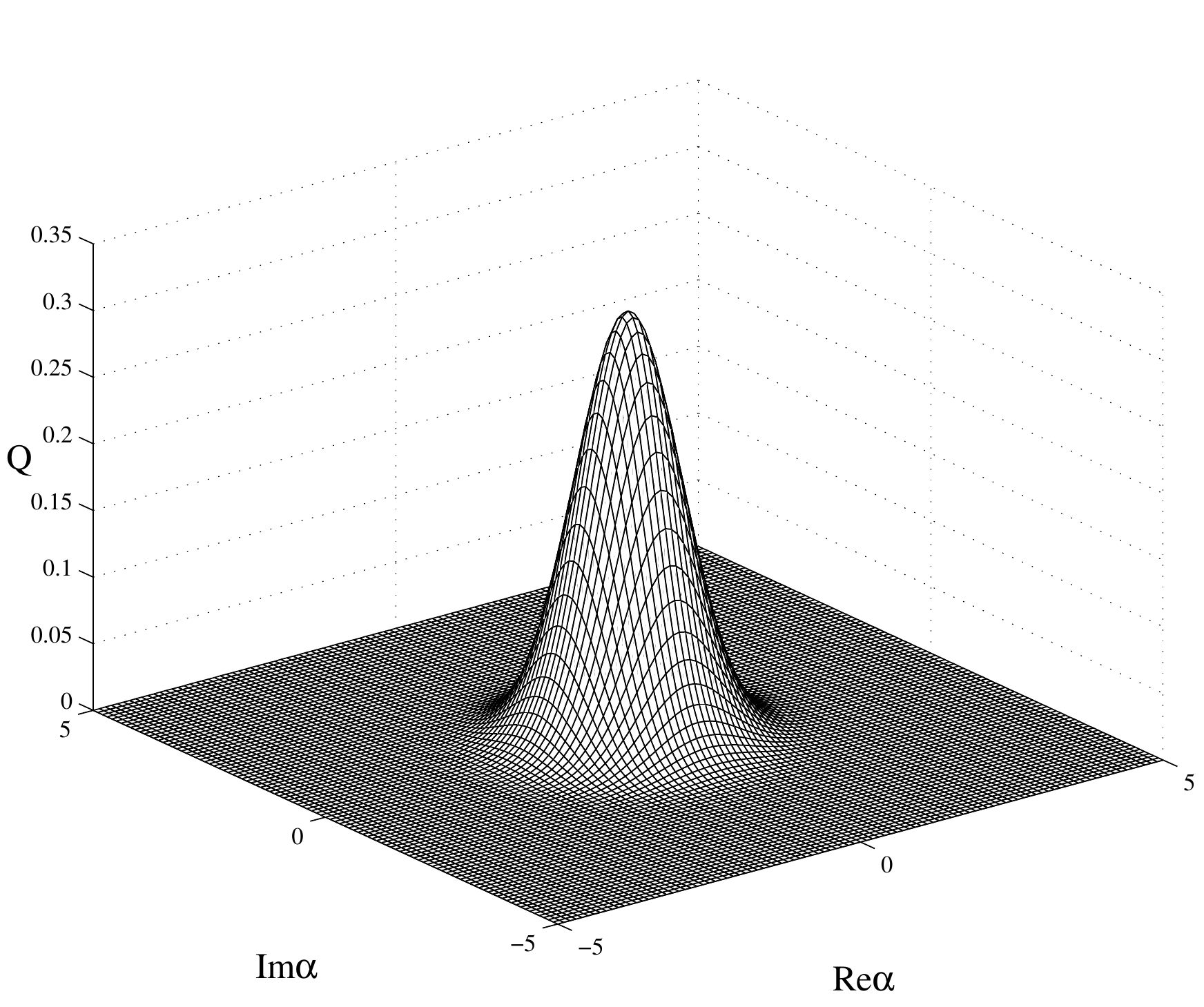} 
\includegraphics[scale=0.35]{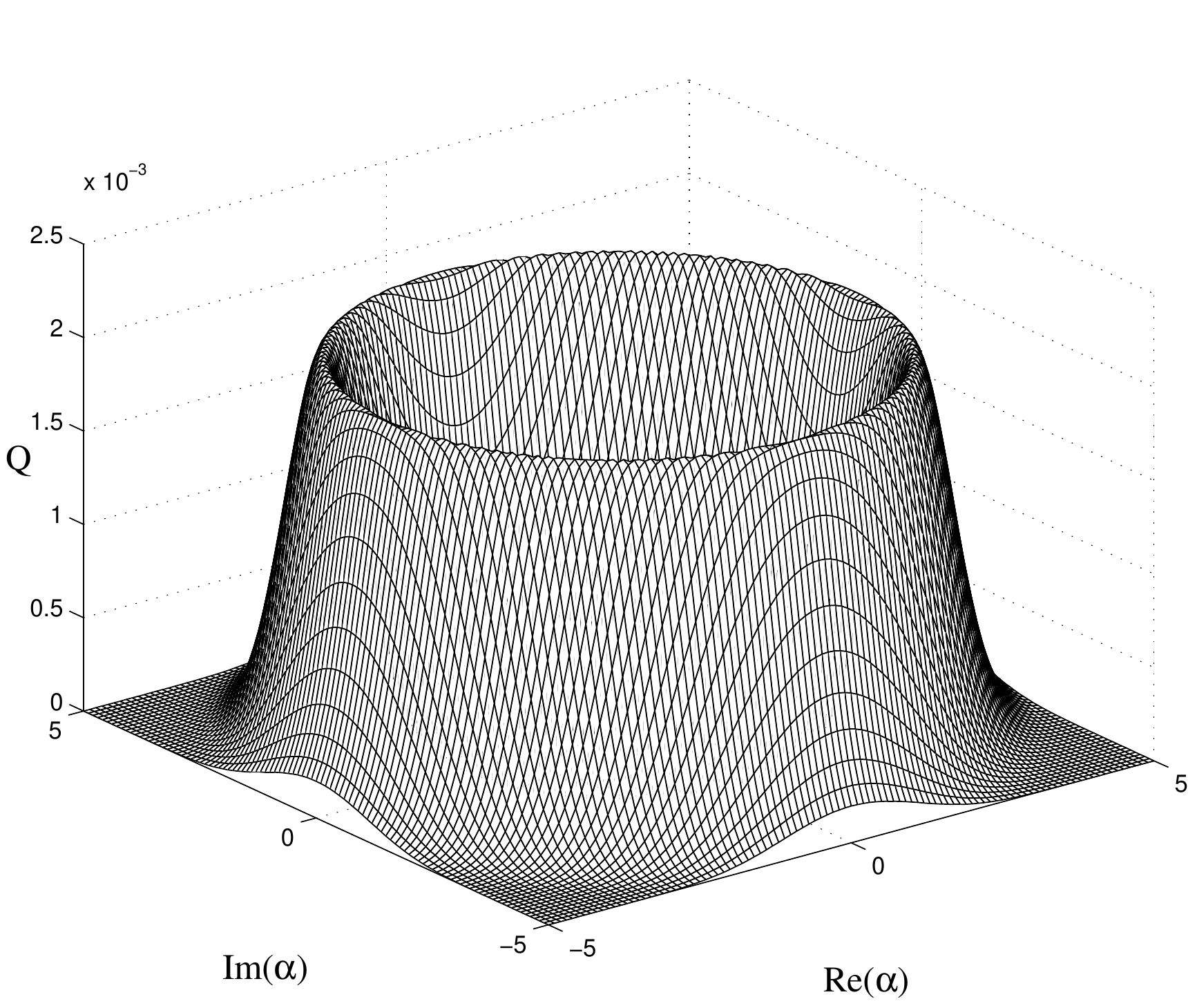} 
\end{center}
\caption{Husimi quasi-probability $Q$-functions for the $n$-photon \textit{Fock} states -- left: $n=0$ (vacuum state), right: $n=4$.}
\label{f1} 
\end{figure}

If we assume that the electric field of the amplitude $\varepsilon$ of the electromagnetic (EM) wave  propagating along the $z$-axis inside the resonator in the form of the planar wave is polarized parallelly to the $x$-axis, it can be described by the following operator
\begin{equation}
\hat{E}_x(z,t)\,=\,\varepsilon\,(\hat{a}+\hat{a}^\dagger )\sin kz
\label{eq5}
\end{equation}
describing electric field of the standing EM wave.
For such a situation and for EM field being in the $n$-photon state, the mean value of the electric field equals zero $\langle n|\hat{E}_x|n\rangle=0$. However the mean value of the squared electric field differs form zero. It can be shown that 
\begin{equation}
\langle n|\hat{E}^2_x(z,t)|n\rangle\,=\,\varepsilon^2\langle n|(\hat{a}^2+(\hat{a}^\dagger)^2+\hat{n}+1)|n\rangle\,=\,
2\varepsilon^2\left(
n+\frac{1}{2}
\right)\,\sin^2kz\quad .
\label{eq6}
\end{equation}
Since the mean value of the electric field equals zero, its squared dispersion $(\Delta E)^2$ can be described by the above equation (eq.\ref{eq6}). It describes the field's fluctuations for the $n$-photon state. Therefore, we can write \citep*{Twww}
\begin{equation}
\Delta E\,=\,\sqrt{\langle E^2\rangle-\langle E\rangle^2}\,=\,
\sqrt{2}\varepsilon\sqrt{n+\frac{1}{2}}\,|\sin kz|\quad .
\end{equation}
What is important, that even for the vacuum state $|0\rangle$ such fluctuations are present and are called \textit{vacuum fluctuations}. In consequence, for $n=0$ we get
\begin{equation}
\Delta E_{\text{vac}}=\,\varepsilon\,|\sin kz|\quad .
\end{equation}
The $n$-photon \textit{Fock} states have a well defined energy (number of photons) whereas their phase is completely undefined. It can be seen when we look at the Husimi $Q$-function plot (Fig.\ref{f1}) where its ring-type structure is visible. Husimi quasi-probabiliy $Q$-function is defined as matrix elements of the density matrix in terms of the \textit{coherent state} $|\alpha\rangle$ in the following way \citep*{H40}
\begin{equation}
Q(\alpha)\,=\,\frac{1}{\pi}\,\langle\alpha|\hat{\rho}|\alpha\rangle\quad .
\end{equation}
For this case the quasi-probability $Q$-function is represented by the following distribution
\begin{equation}
Q(\alpha )\,=\,\frac{1}{\pi}\,\frac{|\alpha |^{2n}}{n!}\,\exp(-|\alpha |^2)\quad .
\end{equation}
Thus, Fig.\ref{f1} shows the plots of Q-function for $n=0$ (vacuum state) and $n=4$ on the complex plane defined by the parameter $\alpha$. It can be seen that for $n=0$ this function exhibits its symmetric one-peak form centered at $\alpha =0$. If the value of $n$ increases (for instance, $n=4$ for Fig.\ref{f1} - right), instead of a single peak the ring-like structure is visible. The radius of this ring is equal to $n$. From these plots it can be seen that, indeed, $n$-photon states have completely undetermined phase, contrary to the energy of such states which is defined very well.

Analogously we can define multi-mode $n$-photon states using individual operators corresponding to the various modes of the field. For instance, the creation and annihilation operators ($\hat{a}^\dagger_k$ and $\hat{a}_k$, respectively) defined for the $k$-\textit{th} mode act on the $f$-mode states as follows \citep*{GK05,Twww}
\begin{equation}
\begin{split}
\hat{a}^\dagger_k\,|n_1,n_2,\ldots ,n_k,\ldots ,n_f\rangle\,=\,&
\sqrt{n_k+1}\,|n_1,n_2,\ldots ,n_k+1,\ldots ,n_f\rangle\\\
\hat{a}_k\,|n_1,n_2,\ldots ,n_k,\ldots ,n_f\rangle\,=\,&
\sqrt{n_k}\,|n_1,n_2,\ldots ,n_k-1,\ldots ,n_f\rangle\quad ,
\end{split}
\end{equation}
where the $f$-mode state can be denoted as
\begin{equation}
|n_1,n_2,\ldots ,n_f\rangle\,=\,
|n_1\rangle |n_2\rangle,\ldots ,|n_f\rangle
\end{equation}
or
\begin{equation}
|n_1,n_2,\ldots ,n_f\rangle\,=\,
|\{n_j\}\rangle\quad .
\end{equation}
For the last case $\{n_j\}$ denotes a set of numbers of photons in every mode.

In particular, the multi-mode vacuum state can be written as
\begin{equation}
|\{0\}\rangle\,=\,|0_1,0_2,\ldots,0_f\rangle
\end{equation}
whereas every state $|\{n_j\}\rangle$ corresponding to the given number of photons in each mode can be obtained from the vacuum state according to
\begin{equation}
|\{n_j\}\rangle\,=\,\prod_j\,\frac{(\hat{a_j}^\dagger)^{n_j}}{\sqrt{n_j!}}\,|\{0\}\rangle\quad .
\end{equation}

\subsection{Finite-dimensional coherent states}
Coherent states, sometimes called as \textit{Glauber} coherent states are "the most classical" states \citep*{GK05} of quantum harmonic oscillator. They can be defined as eigen-states of the annihilation operator $\hat{a}$
\begin{equation}
\hat{a}\,|\alpha\rangle\,=\,\alpha\,|\alpha\rangle\quad ,
\end{equation}
where $\alpha$ is a complex number. This parameter is related to the classical electric field strength. In fact, the value $|\alpha |^2$  equals the mean number of photons $\langle \hat{n}\rangle$ in the field described by the state $|\alpha\rangle$.

The coherent state can be expressed in the $n$-photon states basis as \citep*{G63a,G63c}
\begin{equation}
|\alpha\rangle\,=\,\exp (-\frac{|\alpha |^2}{2})\,\sum_{n=0}^\infty\,
\frac{\alpha^n}{\sqrt{n!}}\,|\alpha\rangle\quad ,
\label{eq17}
\end{equation}
and in consequence, the probability of the observation of $n$ photons  equals
\begin{equation}
P_n\,=\,|\langle n|\alpha\rangle|^2\,=\,\exp (-|\alpha |^2)\,
\frac{\alpha^{2n}}{n!}\,=\,
\exp (-\langle \hat{n}\rangle)\,
\frac{\langle \hat{n}\rangle^{n}}{n!}
\quad .
\label{eq18}
\end{equation}
This probability distribution is a \textit{Poisson} distribution for a mean number of photons  equals $\langle\hat{n}\rangle = |\alpha |^2$ and for high values of $\langle\hat{n}\rangle$ it can be approximated by the Gaussian one.  Moreover, coherent states are not orthogonal in general, i.e. $|\langle\beta |\alpha\rangle|^2\,=\,\exp\big[-|\beta-\alpha|^2\big]$,
however those states can be treated as almost orthogonal for large values of $|\alpha-\beta|$.

\begin{figure}[h!]
\begin{center}
\includegraphics[scale=0.5]{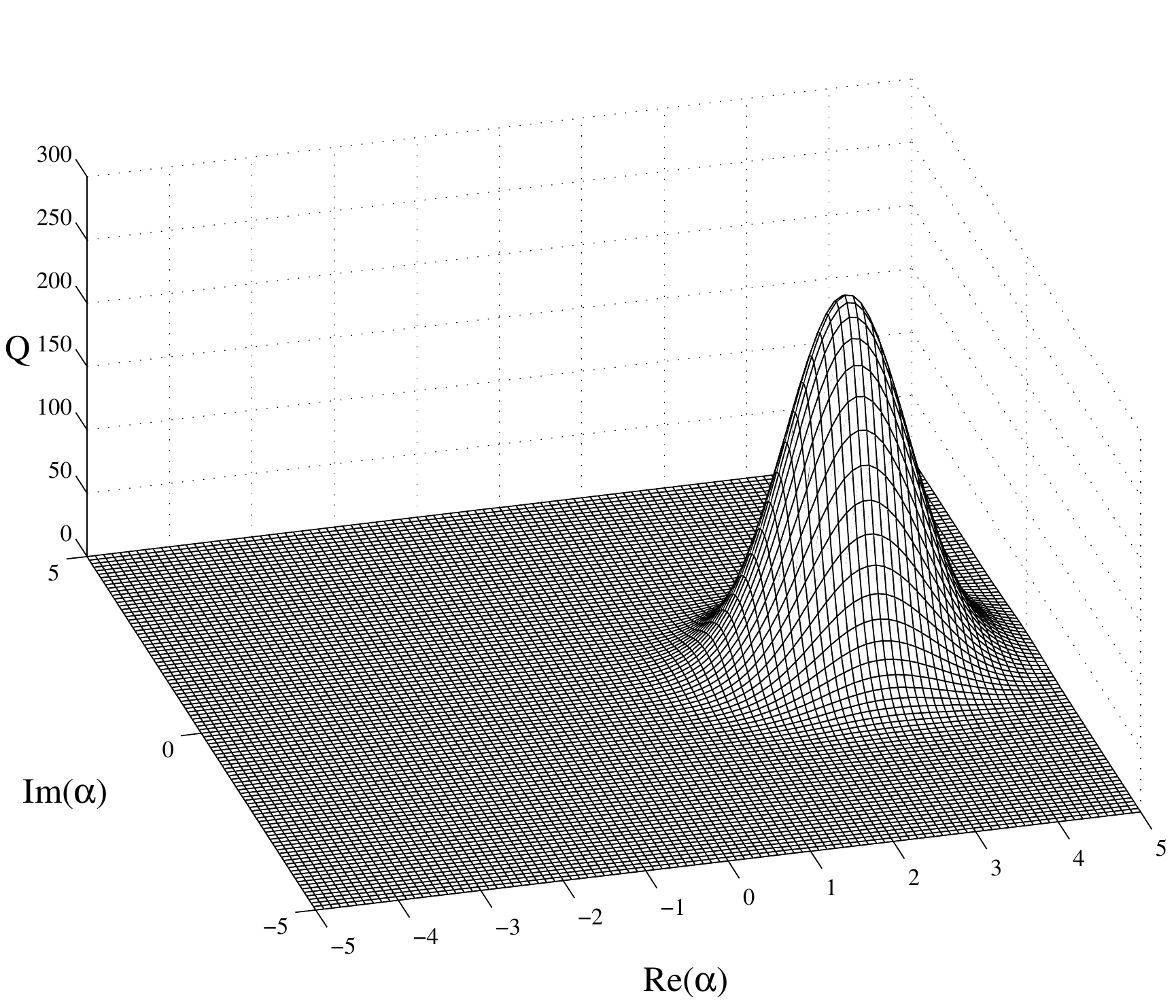} 
\end{center}
\caption{$Q$-function for the coherent state with the mean number of photons $\langle\hat{n}\rangle=|\alpha |^2=3$}.
\label{f2}
\end{figure}

The fluctuations of these states are the same as for the vacuum state and the $Q$-function is of the same shape as that for the vacuum state. However, the maxima of these functions are located at different positions on the complex plane $\alpha$. For the coherent state the center of the peak is shifted form the point $\alpha =0$ to another one, which is distant from $0$ by the value of the mean number of photons (see Fig.\ref{f2}). Therefore, the coherent state can be treated as a \textit{shifted vacuum state}. This feature was applied in the Glauber definition of the coherent state \citep*{G63c} where a displacement operator was used. This definition is equivalent to the formula
\begin{equation}
|\alpha\rangle\,=\,\hat{D}(\alpha,\alpha^*)\,|0\rangle\quad ,
\label{eq19}
\end{equation}
where the displacement operator $\hat{D}(\alpha,\alpha^*)$ is expressed as
\begin{equation}
\hat{D}(\alpha,\alpha^*)\,=\,e^{\alpha\hat{a}^\dagger-\alpha^*\hat{a}}\quad .
\label{eq20}
\end{equation}
This definition can be applied to the \textit{finite dimensional coherent states} (FDCS) construction. Thus, Bu\v{z}ek \textit{et al.} \citep*{BWK92}, Miranowicz \textit{et al.} \citep*{MPT94} and Opatrny \textit{et al.} \citep*{OMB96} proposed a definition analogous to that of Glauber's, but in their definitions the annihilation and creation operators were defined in a truncated Hilbert space. Their proposals were similar to the idea applied in the phase operators formalism proposed by Pegg and Barnett \citep*{PB88,PB89}. They defined operators  in the $(s+1)$-dimensional Hilbert space and then calculated all physical quantities within this space. Just after performing these calculations the limit $s\rightarrow \infty$ was taken. Obviously, it should be kept in mind that for the cases discussed in this papers, this limit is not of interest.

As it was defined in \citep*{BWK92}, for the $s$-dimensional Hilbert space creation and annihilation operators
$\hat{a}^\dagger_{(s)}$ and $\hat{a}_{(s)}$, respectively can be rewritten in terms of projection operators as
\begin{equation}
\begin{split}
\hat{a}^\dagger_{(s)}\,&=\,\sum^s_{n=1}\sqrt{n}\,|n\rangle\langle n-1|\\
\hat{a}_{(s)}\,&=\,\sum^s_{n=1}\sqrt{n}\,|n-1\rangle\langle n|\quad ,
\label{eq21}
\end{split}
\end{equation}
where the number states $|n\rangle$ are defined for the harmonic oscillator in finite-dimensional ($(s+1)$-dimensional) space. These states obey the following relations
\begin{equation}
\langle n|m\rangle\,=\delta_{n,m}\qquad\text{and}\qquad
\sum_{n=0}^s\,|n\rangle\langle n|\,=\,\hat{I}\quad 
\label{eq22}
\end{equation}
and the actions of the operators $\hat{a}^\dagger_{(s)}$ and $\hat{a}_{(s)}$ on the $n$-photon states are defined by relations
\begin{eqnarray}
\hat{a}^\dagger_{(s)}\,|n\rangle =&\,\sqrt{n+1}\,|n+1\rangle
\, ,\qquad &\hat{a}^\dagger_{(s)}\,|s\rangle =0\nonumber\\
\hat{a}_{(s)}\,|n\rangle =&\,\sqrt{n}\,|n-1\rangle
\, ,\qquad &\hat{a}_{(s)}\,|0\rangle =0\quad .
\label{eq23}
\end{eqnarray}
It should be noted that the commutation relation for such defined operators equals
\begin{equation}
[\hat{a}_{(s)},\hat{a}^\dagger_{(s)}]\,=\,1-(s+1)\,|s\rangle\langle s|\quad ,
\label{eq24}
\end{equation}
whereas the number operator $\hat{n}_{(s)}$ is given by the usual formula
\begin{equation}
\hat{n}_{(s)}\,=\,\sum_{n=1}^s\,|n\rangle\langle n|\quad .
\label{eq25}
\end{equation}
It should be pointed that according to the notation proposed in \citep*{BWK92}, the indices used in the sums within the above definitions changes their values from one, not from zero.

An alternative definition of the coherent states in finite-dimensional space  was proposed by Kuang \textit{et al.} \citep*{KWZ93,KWZ94}. They  defined the coherent states by the application of the expansion of the coherent state in a number states basis, analogously as in eq.(\ref{eq17}), but with the finite upper limit of summation, i.e.
\begin{equation}
|\tilde{\alpha}\rangle_{(s)}\,=\,\mathcal{N}_{(s)}\,\sum_{n=0}^s\,
\frac{\alpha^n}{\sqrt{n!}}\,|\alpha\rangle\quad ,
\label{eq26}
\end{equation}
where the normalization coefficient $\mathcal{N}_{(s)}$ can be expressed by the generalized Laguerre polynomial $L_s^n(x)$ as
\begin{equation}
\mathcal{N}_{(s)}\,=\,\left(
\sum_{n=0}^s\,\frac{|\tilde{\alpha}|^{2n}}{n!}
\right)^{-1/2}=\,
\left\{
(-1)^2L_s^{-s-1}(|\tilde{\alpha}|^2)
\right\}^{-1/2}\quad .
\label{eq27}
\end{equation}
Such definition is equivalent to that in which the state $|\tilde{\alpha}\rangle_{(s)}$ is generated as a result of the action of the operator $\exp (\tilde{\alpha}\hat{a}^\dagger)$ on the vacuum state $|0\rangle$. These states are often called \textit{truncated coherent states} (TCS) contrary to the finite-dimensional coherent states (FDCS), defined in \citep*{BWK92,MPT94,OMB96}. For sufficiently high values of the parameter $s$ both states become practically identical to the coherent states defined in the infinite dimensional Hilbert space. The properties of the states defined with application of  both here mentioned definitions  have been  discussed in numerous papers. For a list of them see for instance the review paper \citep*{MLI01} \textit{and the references quoted therein}. For instance, periodic and quasi-periodic properties of  FDCS and TCS were discussed in \citep*{LM96,LMT97}, whereas the  Wigner functions, corresponding to these states, were considered in  \citep*{MLI01}.

Considerations concerning harmonic oscillator in a finite-dimensional Hilbert space lead also to the \textit{odd} and \textit{even} FDCS concepts. These states were considered in \citep*{RR98}, where the following definitions were proposed on the basis of coherent spin states idea \citep*{R71}
\begin{equation}
|\mu\rangle_e\,=\,N_e\cosh (\mu \hat{A}^\dagger)\,|0\rangle\qquad\text{(even states)}
\end{equation}
and
\begin{equation}
|\mu\rangle_a\,=\,N_o\sinh (\mu \hat{A}^\dagger)\,|0\rangle\qquad\text{(odd states)}\quad .
\end{equation}
The operators $\hat{A}^\dagger$ and $\hat{A}$ can be interpreted as the creation and annihilation operators satisfying algebra related to the SU(2) algebra \citep*{RR98}, $\mu$ is a complex number, whereas the parameters $N_e$ and $N_o$ are normalization constants. The states defined in such a way can be expressed in the photon number basis similarly as in eq.(\ref{eq26}) but with \textit{even} or \textit{odd} terms missing.

\subsection{Finite-dimensional squeezed states}
The next group of states commonly discussed in quantum-optical models are those referred to as \textit{squeezed states}. This paper concentrates on the quadrature squeezed states. These states have been extensively discussed in the literature (for instance see \citep*{P84,GK05,Twww} \textit{and the references quoted therein}) and only some basic information concerning them will be given.

The nature of these states is related to the basic ideas of quantum mechanics. Thus, if we have three operators $\hat{A}$, $\hat{B}$ and $\hat{C}$ obeying the commutation relation $[\hat{A},\hat{B}]=i\hat{C}$, the following inequality is valid
\begin{equation}
\langle(\Delta\hat{A})^2\rangle\langle(\Delta\hat{B})^2\rangle\ge
\frac{1}{4}|\langle\hat{C}\rangle|^2\quad
\label{eq30}
\end{equation}
where the variances $\Delta \hat{Y}$ for the given operator $\hat{Y}$ are defined by
\begin{equation}
\Delta\hat{Y}\,=\,\sqrt{\langle\hat{Y}^2\rangle-\langle\hat{Y}\rangle^2}
\end{equation}
and the expectation values of the operators are calculated for a given state $|\psi\rangle$ as $\langle\hat{Y}\rangle\,=\,\langle\psi|\hat{Y}|\psi\rangle$. Thus, the state $|\psi\rangle$ is a \textit{squeezed state} if 
\begin{equation}
\langle(\Delta\hat{A})^2\rangle\,<\,\frac{1}{2}\,|\langle\hat{C}\rangle|^2\quad\text{or}\quad\langle(\Delta\hat{B})^2\rangle\,<\,\frac{1}{2}\,|\langle\hat{C}\rangle|^2\quad .
\end{equation}
Moreover, the state $|\psi\rangle$ is called intelligent if (\ref{eq30}) becomes strict equality (see for instance \citep*{L90} or \citep*{MG05} and \textit{the references quoted therein}). 

For quantum optical fields we can define the \textit{quadrature operators}  $\hat{X}_{1,2}$ using boson creation and annihilation ones
\begin{equation}
\begin{split}
\hat{X}_1\,&=\,\frac{1}{2}\,(\hat{a}+\hat{a}^\dagger)\\
\hat{X}_2\,&=\,\frac{1}{2i}\,(\hat{a}-\hat{a}^\dagger)\quad .
\end{split}
\end{equation}
These \textit{quadrature operators} obey the commutation rule $[\hat{X}_1,\hat{X}_2]=\frac{i}{2}$ and hence, the following uncertainty relation
\begin{equation}
\Delta\hat{X}_1\,\Delta\hat{X}_2\,\ge\,\frac{1}{4}\quad .
\end{equation}
It should be noted that for the coherent state $|\alpha\rangle$ the mean values of the quadratures are
\begin{equation}
\langle\alpha|\hat{X}_1|\alpha\rangle\,=\,\text{Re}\,\alpha\quad\text{and}\quad
\langle\alpha|\hat{X}_2|\alpha\rangle\,=\,\text{Im}\,\alpha
\label{e35}
\end{equation}
and their uncertainties equal $1/4$ for the both operators (they equal  those for the vacuum state). It is possible to decrease the uncertainty in one of the quadratures simultaneously increasing the second one. In consequence, one of them can be smaller than $1/4$. Such states are referred to as \textit{squeezed states} of light \citep*{W83}. 

Sometimes we use \textit{normally ordered} variances $\langle :(\Delta\hat{X})^2:\rangle$ (where for instance $ :\hat{a}\hat{a}^\dagger:=\hat{a}^\dagger\hat{a}$) that both equal zero for the coherent states. If the optical field is in a \textit{squeezed state}, such a variance corresponding to the one of quadratures is negative. 

\textit{Squeezed state} can be defined as a result of the action of the \textit{squeezing} operator on the coherent state. On the other hand, the latter can be generated by the action of the \textit{dispalcement} operator on the vacuum state \citep*{Twww}. Thus, it can be writen
\begin{equation}
|\alpha ,\xi\rangle\,=\,\hat{D}(\alpha)\,\hat{S}(\xi)\,|0\rangle
\end{equation}
where the displacement operator $\hat{D}(\alpha)$ (eq.(\ref{eq20})) have already been discussed in the context of coherent state, whereas \textit{squeezing} operator is defined as
\begin{equation}
\hat{S}(\xi)\,=\,e^{\frac{1}{2}(\xi^*\hat{a}^2-\xi\hat{a}^{\dagger 2})}\quad ,
\label{eq37}
\end{equation}
where the \textit{squeezing} parameter $\xi$ is a complex number.

\textit{Squeezed} state can be expressed in terms of number states $|n\rangle$. The probability amplitude corresponding to the $n$-photon state for the state $|\alpha,\xi\rangle$ is of more complicated form as we compare it with that for the coherent state and can be expressed in terms of the Hermite polynomial $H_n$ of degree $n$. Thus, we can write
\begin{equation}
|\alpha,\xi\rangle\,=\,\sum_{n=0}^\infty\,c_n\,|n\rangle
\end{equation}
where
\begin{equation}
\begin{split}
c_n\,=\,\frac{1}{\sqrt{n!\,\cosh r}}\,\left[
\frac{1}{2}\,\exp (i\theta)\,\tanh r
\right]^{n/2}\,H_n\left(
\frac{\alpha+\alpha^*\exp (i\theta)\tanh r}{\sqrt{2\exp (i\theta) \tanh r}}
\right)&\\
\times\,\exp\left[
-\,\frac{1}{2}\,(|\alpha |^2+(\alpha^*)^2\exp (i\theta)\,\tanh r)
\right]&\quad 
\end{split}
\end{equation}
and the complex parameters are expressed in the form $\alpha=|\alpha|\exp (i\phi)$ and $\xi=|\xi|\exp (i\theta)$. For a special case, when $\alpha =0$ our state becomes \textit{squeezed vacuum} state $|0,\xi\rangle=\hat{S}(\xi)\,|0\rangle$ and we can write for the probability amplitudes 
\begin{equation}
c_{2n}\,=\,\frac{1}{\sqrt{\cosh r}}\,\frac{\sqrt{(2n)!}}{n!}\,\left[
-\,\frac{1}{2}\,\exp (i\theta)\,\tanh r
\right]^n
\end{equation}
for even numbers of photons. For odd numbers this amplitude equals zero – it is not possible to find odd numbers of photons for such a state.

\begin{figure}[h!]
\includegraphics[scale=0.4]{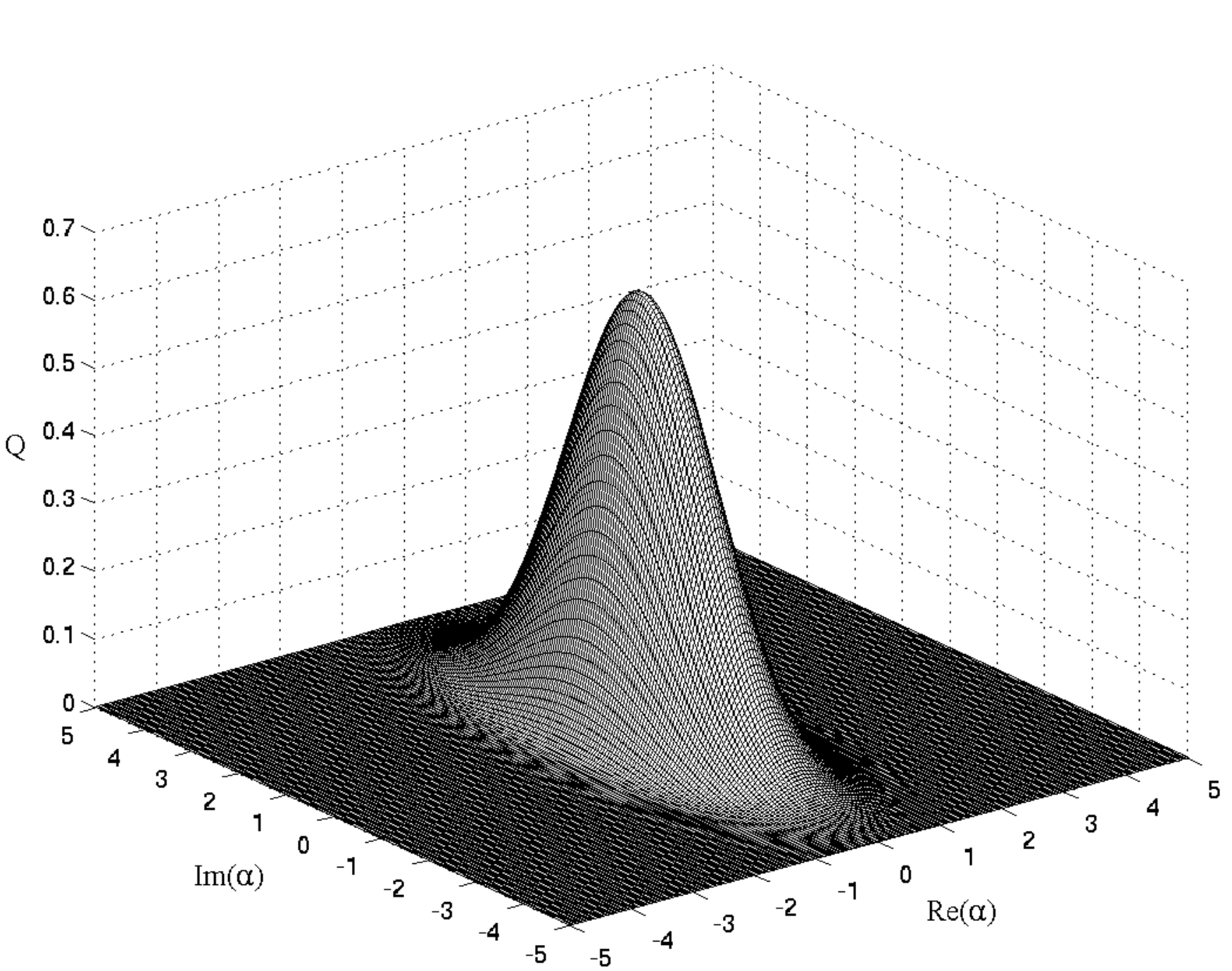} 
\includegraphics[scale=0.4]{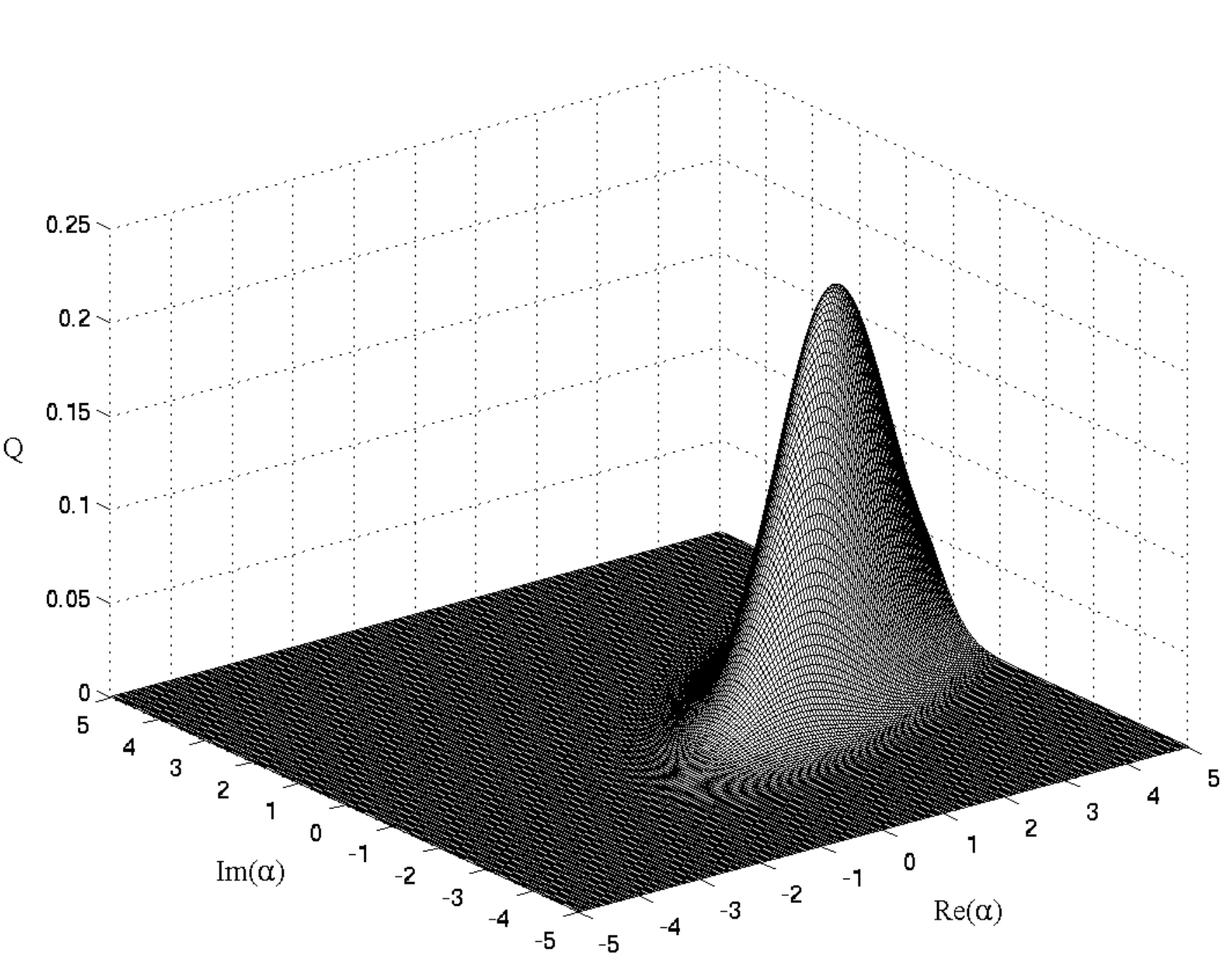} 
\caption{$Q$-function plots for the squeezed vacuum (left) and squeezed state with $|\alpha|=3$ (right). Squeezing parameters are equal to $i$ and $-\frac{1}{2}-i\frac{\sqrt{3}}{2}$, respectively.}
\label{f3} 
\end{figure}

The form of the Husimi $Q$-function reflects the properties of the \textit{squeezed} states. From its plot it is possible to determine how far and in which direction the state is shifted in the complex phase plane, and in which direction and how much is squeezed. For instance, Fig.\ref{f3} shows $Q$-function for the squeezing angle $\theta =\pi/4$ and two values of $|\alpha|$. For $|\alpha|=0$ we have $Q$-function for the squeezed vacuum state, whereas for $|\alpha|=3$ the plot is shifted along the Re$(\alpha)$ axis.

Such defined \textit{squeezed states} can be a starting point for \textit{finite-dimensional squeezed states} (FDSS) and \textit{finite dimensional squeezed vacuum} (FDSV) definitions. These states can be defined the analogous way as previously discussed FDCS. Moreover,  we can define \textit{truncated squeezed vacuum} (TSV) and \textit{truncated squeezed states} (TSS) following the path shown for TCS and definition. 

Now, without losing the generality of our considerations, we shall restrict ourselves to the presentation of FDSV and TSV states only. Of course, applying the methods presented here it is possible to extend the results obtained for FDSV (TSV) to the cases of FDSS (TSS).

Thus, using the creation $\hat{a}_{(s)}^\dagger$ and annihilation $\hat{a}_{(s)}$  operators defined in finite dimensional Hilbert space  squeezing operator can be redefined (see eqns.(\ref{eq21},\ref{eq37})) and acting it on the vacuum state we get FDSV \citep*{MLDT96,MLT98,MLI01}. The state generated in such a way can be expressed in the $n$-photon  basis as
\begin{equation}
|0,\xi\rangle_{(s)}\,=\, \sum_{n=0}^\sigma\,b_{2n}^{(s)}\,\exp (in\phi)\,|2n\rangle\quad ,
\end{equation}
where the probability amplitude $b_{2n}^{(s)}$ is given by
\begin{equation}
b_{2n}^{(s)}\,=\,(-i)^n\frac{(2\sigma)!}{\sqrt{(2n)!}}\,\sum_{k=0}^\sigma\,
\exp (i|\xi |x_k/2)\,\frac{G_n(x_k)}{G_\sigma (x_k)G_{\sigma +1}'(x_k)}\quad .
\end{equation}
The parameter $\sigma=[\![ s/2]\!]$ whereas $G_n(x)$ are the Meixner-Sheffer orthogonal polynomials. They can be defined by the following relation
\begin{equation}
G_{n+1}\,=\,xG_n-2n\,(2n-1)\,G_{n-1}\quad\text{for}\quad n=2,3,\ldots
\end{equation}
with $G_0(x)=1$ and $G_1(x)=x$. Moreover, $x_k\equiv x_k^{(\sigma +1)}$ is the $k$-th root ($k=0,\ldots ,\sigma$) of the polynomial $G_{\sigma +1}(x)$ and $G_{\sigma +1}'$ is the $x$ derivative at $x=x_k$.

On the other hand, one can define TSV by truncation of $n$-photon expansion of the squeezed vacuum defined in infinite dimensional Hilbert space and, after it, performing the proper normalization. As a result TSV can be defined as \citep*{MLI01}
\begin{equation}
|\bar{\xi}\rangle\,=\,\sum_{n=0}^{\sigma +1}\,
b_{2n}^{(s)}\,\exp\,(in\phi)\,|2n\rangle
\end{equation}
where the coefficient $b_{2n}^{(s)}$ is given by
\begin{equation}
b_{2n}^{(s)}\,=\,\mathcal{N}_s\,\frac{\sqrt{(2n)!}}{4n!}\,\tanh^n|\bar{\xi}|
\end{equation}
with the normalization parameter $\mathcal(N)_s$ defined by the formula
\begin{equation}
\mathcal(N)_s^{-2}\,=\,\cosh |\bar{\xi}|-2t^{2\sigma+2}
\binom{2\sigma +1}{\sigma}
\,_2F_1(1,\{3/2+\sigma,\,2+\sigma\},\,4t^2)\quad .
\end{equation}
The parameter $\sigma=[\![s/2]\!]$ whereas the function $_2F_1$ is the generalized hyper-geometrical function. The squeeze parameter $\bar{\xi}$ and the corresponding to it state $|\bar{\xi}\rangle$ were marked by the bar to distinguish them from their counterparts defined for FDSV.

\subsection{Multi-mode states -- Bell states}
Up to this point we were interested in single-mode states defined in the finite-dimensional Hilbert space and we have presented here a short description of them showing some basic facts concerning their definitions, nature and properties. The next natural step is an extension of our considerations towards the multi-mode states. In practice, such states can be defined on the basis of every kind of the already discussed states. Even for the short list of the presented here single-mode states many possible cases of the multi-mode states and their definitions can be expected. Therefore, since this paper is devoted to the methods of the finite-dimensional states generation, not to their properties thorough description, we shall concentrate on the Bell (and generalized Bell), GHZ (Greenberg-Horne-Zeilinger) and $W$ states only. These states are defined with the application of the $n$-photon \textit{Fock} states which can be easily manipulated from the mathematical point of view, and are widely discussed in many problems related to the quantum information theory.

The presentation of those states starts from the simplest case of the Bell states. The idea of these states is a one of the basic quantum information theory concepts named after the \textit{John F. Bell} and is related to his famous inequality \citep*{B64}. The Bell's inequality was developed by Clauser \textit{et.al} in \citep*{CHS69} where the proposal of the experimental verification of the \textit{local hidden-variable theory} was presented -- see also \citep*{CH74}. Therefore, this inequality is often referred as to the \textit{Bell--CHSH inequality} or CHSH one. This inequality derivation was an important point in the discussion of the \textit{Einstein-Podolsky-Rosen} paradox \citep*{EPR35} and its consequences in the understanding of the present quantum mechanics's theory. Due to the relations of the Bell states with the EPR paradox they are sometimes called \textit{EPR pair}. For the description of the Bell states, the discussion concerning their properties and applications see for example \citep*{NH00,H06,LP07,D07,J07} \textit{and the references quoted therein}).  

Thus, if we restrict our considerations to the case when only two states in each mode are involved - the vacuum state and one-photon state, the Bell states can be expressed in the following form
\begin{equation}
\begin{split}
|\psi_-\rangle\,=&\,\frac{1}{\sqrt{2}}\,\biggl(
|0\rangle_A|1\rangle_B\,-\,|1\rangle_A|0\rangle_B
\biggr)\\
|\psi_+\rangle\,=&\,\frac{1}{\sqrt{2}}\,\biggl(
|0\rangle_A|1\rangle_B\,+\,|1\rangle_A|0\rangle_B
\biggr)\\
|\phi_-\rangle\,=&\,\frac{1}{\sqrt{2}}\,\biggl(
|0\rangle_A|0\rangle_B\,-\,|1\rangle_A|1\rangle_B
\biggr)\\
|\phi_+\rangle\,=&\,\frac{1}{\sqrt{2}}\,\biggl(
|0\rangle_A|0\rangle_B\,+\,|1\rangle_A|1\rangle_B
\biggr)\quad ,
\label{eq47}
\end{split}
\end{equation}
where we assume that for each of the two modes (labeled as $A$ or $B$) we have $0$ and/or $1$ photon. These states are examples of the two-\textit{qubits} states which are \textit{maximally entangled states}. If the quantum system is in one of these states, the Bell--CHSH inequality leads to the correlation measure equal to $2\sqrt{2}$ instead of $2$ predicted by the local theories. 

We can expand the idea of the Bell states to the case when there are more photons in the system -- when we deal with $|n\rangle$ state instead of the one-photon state $|1\rangle$. For this situation we have \textit{NOON}-states that have been proposed by Sanders \citep*{S89} in the context of the quantum dynamics of the nonlinear rotator. Such states play an important role in the quantum theory, for instance in quantum metrology problems. \textit{NOON}-states can be expressed as
\begin{equation}
|\psi\rangle_{NOON}\,=\,\frac{1}{\sqrt{2}}\,\biggl(
|N\rangle_A|0\rangle_A\,+\,\exp (iN\theta)|0\rangle_A|N\rangle_B
\biggr)\quad ,
\label{eq48}
\end{equation}
where the parameter $\theta$ denotes some quantum phase. 

The idea of the Bell states was also extended to the \textit{generalized Bell states}. In \citep*{BBC93} quantum teleportation via dual (classical and quantum) channels was discussed and the following states were defined. These states can be expressed as in \citep*{SL09}
\begin{equation}
|\Psi_{mn}\rangle=\frac{1}{\sqrt{D}}\sum_{k=0}^{D-1}e^{2\pi i k m/D}|k\rangle\otimes|(k-n)\mbox{mod} D\rangle\quad .
\label{eq49}
\end{equation}
When we restrict our consideration to the case when we have only the vacuum, one-photon and two-photons states in the system, some of the generalized Bell states become
\begin{equation}
\begin{split}
|B_1\rangle\,&=\,\frac{1}{\sqrt{3}}\,\left(
|0\rangle_a|0\rangle_b\,+\,|1\rangle_a|1\rangle_b\,+\,|2\rangle_a|2\rangle_b
\right)\\
|B_2\rangle\,&=\,\frac{1}{\sqrt{3}}\,\left(
|0\rangle_a|0\rangle_b\,+\,e^{i\frac{2\pi}{3}}\,|1\rangle_a|1\rangle_b\,+\,e^{-i\frac{2\pi}{3}}\,|2\rangle_a|2\rangle_b
\right)\\
|B_3\rangle\,&=\,\frac{1}{\sqrt{3}}\,\left(
|0\rangle_a|0\rangle_b\,+\,e^{-i\frac{2\pi}{3}}\,|1\rangle_a|1\rangle_b\,+\,e^{i\frac{2\pi}{3}}\,|2\rangle_a|2\rangle_b
\right)\quad .
\end{split}
\label{eq50}
\end{equation}

An extension of the idea of Bell states can lead to other quantum states that are widely discussed in the quantum information theory. One of them are \textit{GHZ} (Greenberg-Horne-Zeilinger) states \citep*{GHZ89}. These states were applied as a tool of the EPR paradox investigation in the models involving three spin-$1/2$ particles \citep*{M90} or those involving multi-particle interferometry  \citep*{GHS90}. The latter model was investigated experimentally in \citep*{PBD00}. 
GHZ states are commonly used for the investigation of the multipartite entanglement in various models of quantum information theory.

GHZ states are entangled states which can be written in general as
\begin{equation}
|\psi\rangle_{n GHZ}\,=\,\frac{1}{\sqrt{2}}\,\biggl(|0\rangle^{\otimes n}\,+\,|1\rangle^{\otimes n}
\biggr)\quad .
\label{eq51}
\end{equation}
The parameter $n$ describes the number of \textit{qubits} which are involved. In particular, if there are $3$-\textit{qubit} GHZ state (we have three subsystems/modes labelled by $A$, $B$ and $C$), it becomes
\begin{equation}
|\psi\rangle_{3 GHZ}\,=\,\frac{1}{\sqrt{2}}\biggl(|0\rangle_A|0\rangle_B|0\rangle_C\,+\,|1\rangle_A|1\rangle_B|1\rangle_C
\biggr)\quad .
\label{eq52}
\end{equation}
Even for this simplest case of the GHZ states $|\psi\rangle_{3 GHZ}$ exhibits non-trivial multipartite entanglement. 
At this point one should mention $W$ states which can be expressed for the three-\textit{qubits} case as
\begin{equation}
|\psi\rangle_{3W}\,=\,\frac{1}{\sqrt{3}}\biggl(
|1\rangle_A|0\rangle_B|0\rangle_C\,+\,|0\rangle_A|1\rangle_B|0\rangle_C\,+\,|0\rangle_A|0\rangle_B|1\rangle_C
\biggr)
\label{eq53}
\end{equation}
and can be generalized for $n$ \textit{qubits} as
\begin{equation}
|\psi\rangle_{nW}\,=\,\frac{1}{\sqrt{n}}\biggl(
|1\rangle_1|0\rangle_2|0\rangle_3\ldots |0\rangle_n\,+\,
|0\rangle_1|1\rangle_2|0\rangle_3\ldots |0\rangle_n\,+\ldots +\,
|0\rangle_1|0\rangle_2|0\rangle_3\ldots |1\rangle_n
\biggr)\quad .
\label{eq54}
\end{equation}
They are also states that exhibits multipartite entanglement however $W$ states exhibit a different nature of the entanglement as they are compared with GHZ states. For instance, for the three-\textit{qubits} case, if we trace 
out one of the \textit{qubits} the remaing two are left entangled. Contrary to this fact, if one part of the GHZ state is traced out, the entanglement for the other parts is destroyed. In this sense one can say that they constitute two \textit{complementary} classes of states \citep*{DVC00} that were discussed in various aspects for the quantum theory models.

\section{Linear quantum scissors}

Quantum scissors (QS) are known in literature as a group of methods or physical devices enabling the generation of the finite superpositions of the number states by truncation of the state of the system which is defined in infinite-dimensional Hilbert space. In this way they belong to a wider group of the quantum states engineering methods. Since in this paper we shall concentrate on the states of  EM field, the scissors discussed here are based on the optical methods in which the truncation can be achieved in various ways. Those methods can make use of only linear optical elements and in such situation one can speak about \textit{linear quantum scissors} (LQS) or alternatively, when nonlinear optical media are used for obtaining finite dimensional states one can speak about \textit{nonlinear quantum scissors} (NQS) devices.

\subsection{LQS models based on the beam-splitters}
The first theoretical scheme suggesting the possibility of physical truncation of the optical coherent state was given by Pegg, Philips and Barnett \citep*{PPB98} and then in \citep*{BP99}. 
For the models discussed there the cut optical coherent state $|\alpha\rangle$ was defined in the two-dimensional space, so its expansion in the $n$-photon basis (see eq.\ref{eq17})) was cut to that of a simple form where the superposition of only two number states (vacuum and one-photon) were involved:
\begin{equation}
|\Psi\rangle_{cut}=\gamma_0|0\rangle+\gamma_1|1\rangle \quad 
\label{eq55}
\end{equation}
with the probability amplitudes $\gamma_{0}$ and $\gamma_{1}$. 
The model proposed there required the use of the linear optical elements only without the necessity of using optical cavity.
Their method is based on the projection synthesis and the scheme for obtaining and measuring  FDCS (\ref{eq55}) is depicted in Fig.4.
The device consists of two symmetric beam-splitters ($BS_1$ and $BS_2$), each of them  reflects and transmits incident field with probability of  $0.5$. If we assume that at the $BS_1$ one of the incident fields (the mode denoted by $\hat{a}_1$) is in the one-photon state, while the other one (the mode $\hat{a}_2$) is in the vacuum state, the beam-splitter produces an entangled state of the form:
\begin{equation}
|\Psi\rangle_{ent}=t_1 |1\rangle_{b_1}|0\rangle_{b2}-r_1^{\star}|0\rangle_{b1}|1\rangle_{b2}
\label{eq56}
\end{equation}
where $t_i$ and $r_i$ are transmission and reflection coefficients. For the case of the symmetric beam-splitters discussed here, they obey the relation $|t_i|^2=|r_i|^2=0.5$.
Thus, obtained entangled state (\ref{eq56}) is mixed at the second beam-splitter  with another 
input state $|\Psi\rangle_{in}$, which is assumed to be a coherent Glauber state (\ref{eq17}). 
The result of the action of the beam-splitter $BS_2$ is a state that can be expressed as in  \citep*{OMKI01}:
\begin{eqnarray}
|\Psi\rangle_{(b_1,c_2,c_3)}&=&e^{-\frac{|\alpha|^2}{2}}\sum\limits_{n=0}^{\infty}\sum\limits_{k=0}^{n}\frac{\alpha^n(-r_2^{\star})^k(t_2)^{n-k}}{\sqrt{k!(n-k)!}}\label{eq_4}\\
&\times&\left( t_1\;|1,k,n-k\rangle-r_1^{\star}r_2\;\sqrt{n-k+1}\;|0,k,n-k+1\rangle -r_1^{\star}t_2^{\star}\sqrt{k+1}\;|0,k+1,n-k\rangle\right), \nonumber
\end{eqnarray}
where $t_i$ and $r_i$ $(i=\{1,2\})$ denote the transmittance and reflectance of the beam splitters, respectively.

Both fields coming out from $BS_2$ are then detected. 
The state $|\Psi\rangle_{out}$ depends solely on the results of the measurements done by the detectors.
If detector $D_1$ measures one photon and detector $D_2$ detects no photons, we obtain the state $_{(c_2,c_3)}\langle 1 0|\Psi\rangle_{(b_1,c_2,c_3)}$ and it is indeed the state coming out from the beam-splitter $BS_1$.
It is easy to verify that this is a superposition of zero and one photon state (\ref{eq55}) with normalizing constant.

As it was shown by Villas-B$\hat{o}$as \textit{et al.} \citep*{BGMB01} the same experimental setup can lead to the more general case of the finite dimensional state generation. This state is of the following form: $|\Psi\rangle_{out}=c_0|0\rangle+\cdots +c_N|N\rangle$ where $N=1,2,\cdots$. The only difference between this model and that described above is in the form of the input state of the beam splitter $BS_1$. In order to obtain $N$ dimensional truncated states one has to fed this beam splitter by the state $|N-1\rangle$ in mode $\hat{a}_2$. The coherent state at the input of the beam splitter $BS_2$ remains the same. The state coming out of the $BS_2$ now has to be specified. If one of the detectors counts one photon and the other one registers $N-1$ photons, after performing the projection procedure we obtain the desired $N$-dimensional superposition of the \textit{Fock} states. The specific form of the  state obtained depends on the values of the transmittances of the both beam splitters $BS_1$ and $BS_2$.
As LQS scheme for obtaining $N$-dimensional truncated states demands a presence of $N-1$- photons sources, in \citep*{BGMB01} the use of experimental setup of quantum Fock filter is proposed similarly as in \citep*{AMPS00}.

A different approach for obtaining more complex states than those from eq.(\ref{eq55}) was proposed  by Koniorczyk {\em et al.} \citep*{KKGJ00}. They propose the same experimental setup as in \citep*{PPB98,BP99}, but the state obtained from proposed scheme is cut in such a way that the two-photon state is involved: 
\begin{equation}
\Psi_{out}=\sqrt{N}\left( \gamma_{0}|0\rangle+\gamma_1 |1\rangle +\gamma_2 |2\rangle\right) \quad ,
\label{eq58}
\end{equation}
where $\sqrt{N}=1/\sum_{k=0}^2|\gamma_k|^2$ is a renormalization constant. Such a state
can be achieved when different than used in \citep*{PPB98,BP99} states are introduced to the scissors device via the beam splitter $BS_1$. As the output state is supposed to contain two photons, it is obvious that these two photons have to be inserted into the device . If this ancillary state is $|11\rangle$ then after the action of the beam splitter $BS_2$ on this particular state and onto the coherent states $|\Psi\rangle_{in}$ the measurement is made by the detectors on the modes $\hat{c}_2$ and $\hat{c}_3$. A necessary condition to observe a desired state  $|\Psi\rangle_{out}$ (\ref{eq58}) is the measurement of one photon by the detector $D_1$ and one photon by the detector $D_2$ as well. For this case the coefficients $\gamma_i$ ($i=0,1,2$) are the same as for the TCS.

There are also some requirements that both beam splitters have to meet in order to generate a desired state (\ref{eq58}) correctly.
The optimal condition, therefore depends on the relations between the phase shifts that both field modes (transmitted and reflected) coming out of beam splitters  gain and the values of transmittance and reflectance. In the simplest case all of the phase shifts equal  zero and both beam splitters are the same with the transmittance value equal to 0.21 or 0.79.

There are also other LQS schemes that are useful for obtaining truncated states containing more than the vacuum and one-photon states only. One of them, based on the projection synthesis described by Dakna {\em et al.} \citep*{DCKW99} uses an array of symmetric beam splitters and efficient photon detectors. This method is proposed for obtaining an arbitrary pure quantum state by successive displacement operations on the current state of the system and adding a single photon. This procedure is performed alternately and under the condition that after each displacement operation and single photon adding, no photon is detected by none of the detectors, one can end the whole procedure after $N$ repetitions obtaining the state of the form \citep*{DCKW99}:
\begin{equation}
|\Psi\rangle\sim \hat{D}(\alpha_{N+1})\hat{a}^{\dagger}T^{\hat{n}}\hat{D}(\alpha_{N})\hat{a}^{\dagger}T^{\hat{n}}\hat{D}(\alpha_{N-1})\cdots \hat{a}^{\dagger}T^{\hat{n}}\hat{D}(\alpha_{1})|0\rangle \quad ,
\label{eq59}
\end{equation}
where $T$ is a transmittance of a beam splitter, $\hat{a}^{\dagger}$ is photon creation operator and $\hat{D}(\alpha)$ is a displacement operator given by $\hat{D}(\alpha)=\exp\left( \alpha\hat{a}^{\dagger}-\alpha^{\star}\hat{a}\right) $.
This state (\ref{eq59}) is equivalent form of a finite-dimensional superposition of Fock states given by $\sum_{n=0}^Nc_n|n\rangle$.

The scheme of the experimental setup suitable for creation of a quantum state in a form (\ref{eq56}) proposed in \citep*{DCKW99} is presented in Fig.\ref{f5}. The proposed device  is composed of an array of ''blocks'' (dashed box in the figure) that produces at the first step a displaced vacuum state by mixing a coherent state with the vacuum state at beam splitter $BS_1$. During the second step  one photon is added and under the condition that the detector $D_1$ registers no photons the state that enters the second ''block'' is formed as: $\hat{a}^{\dagger}T^{\hat{n}}\hat{D}(\alpha_{1})|0\rangle$. After action of the $N$ repeated ''blocks'' the state coming out from the setup has a form presented by (\ref{eq59}).

\subsection{LQS based on interferometric setups}

The truncated state composed of the vacuum and a one photon states (\ref{eq55}) in any configuration can be also obtained in a system with specially arranged two Mach-Zehnder interferometers and two avalanche photodetecors as in \citep*{P00}. The scheme of a device is shown in Fig.\ref{f6} where the beam splitters forming Mach-Zehnder interferometers are identical and symmetric. The phase shifts: $\Theta_1$ in the upper interferometer and  $\Theta_2$ in the bottom one can be changed.
 The one-photon state in mode $\hat{b}$ and the vacuum state in mode $\hat{a}$ are the input states for the first of the two interferometers. One of the output states from the first interferometer becomes one of the input states of the second interferometer while the second  output state from the first interferometer forms a desired state $|\Psi\rangle_{out}$. The form of the latter is known after performing an appropriate detection by the photon counters $D_c$ and $D_b$. The amplitudes of these states are determined by the phase shift introduced between the field traveling through both arms of the interferometer.
The other state (mode $\hat{c}$) that enters the second interferometer is a coherent state. The detectors $D_c$ and $D_b$ measure number of photons in modes $\hat{b}$ and $\hat{c}$ at the output of the whole two interferometer systems. Also in this second interferometer a phase shift $\Theta_2$ can be varied and in that way the balance between the amplitudes in modes $\hat{b}$ and $\hat{c}$ is determined.
The action of each interferometer can be written for $n=1,2$ as \citep*{P00}:
\begin{equation}
\hat{U}(\Theta_n/2)=\exp\left[ \frac{i\pi}{2}\,\hat{b}^{\dagger}\hat{b}\right] \exp\left[ -i\frac{\Theta_n}{2}\left( \hat{a}^{\dagger}\hat{b}+\hat{b}^{\dagger}\hat{a}\right) \right] \exp\left[ -\frac{i\pi}
{2}\,\hat{b}^{\dagger}\hat{b}\right]
\label{eq60}
\end{equation}
and the output state is constructed as an action of an operator $\hat{U}(\Theta_n/2)$ on an input state $|\Psi\rangle_{in}$. The measurements done by both photodetectors on appropriate output modes then determine the specific form of the output state in mode $\hat{a}$. In such a way, similarly to other linear quantum scissors devices, the output state is a conditional one since it depends on the measurements outcomes. The conditions for obtaining any superposition of vacuum and one-photon state are described in \citep*{P00} as well. The first one is an appropriate measurement: one photon count registered by one of the detectors and no photons detected by the other one. The output state  has then the following form \citep*{P00}:
\begin{equation}
|\Psi\rangle_{out}=\frac{\sin\frac{\Theta_1}{2}\sin\frac{\Theta_2}{2}|0\rangle+\gamma\cos\frac{\Theta_1}{2}\cos\frac{\Theta_2}{2}|1\rangle}{\sqrt{\sin^2\frac{\Theta_1}{2}\sin^2\frac{\Theta_2}{2}+|\gamma|^2\cos^2\frac{\Theta_1}{2}\cos^2\frac{\Theta_2}{2}}}
\label{eq_61}
\end{equation}
It is easily seen from (\ref{eq_61}) that the values of internal interferometer phase shifts and the amplitude of the input coherent state (which in fact is cut by this interferometric device) determine the particular form of the output state and the probability of detection of the desired superposition state is also determined. In the example given for a specific amplitude of the coherent state, there exist a range of phase shifts that makes detection probability  larger than $20\%$, which is a better result than the one which can be obtained for LQS device presented in Fig.\ref{f5} \citep*{DCKW99}.
An interferometric scheme has an advantage over the devices presented in Figs.\ref{f4} and \ref{f5}. It is more stable and besides using a larger number of optical elements is able to achieve larger detection probabilities of a desired conditional output state than for the devices described earlier.

Another interferometric approach to the LQS devices is done by Miranowicz in \citep*{M05} -- see Figure 7. Further generalization of the truncation procedure enables obtaining higher dimensional \textit{qudits} or selective truncation and is based on an optical multiport interferometry. 
The presented interferometer setup although previously theoretically and experimentally explored has not been used as a quantum scissors device. After proper adaption it can be used for construction of the truncated states.

The scheme proposed in that paper can be used for obtaining up to $6$-dimensional \textit{qudits} including all of the truncated states mentioned in the previous papers \citep*{PPB98,BP99,KKGJ00}. The interferometer consists of a mirror, a sequence of beam splitters and phase shifters arranged in such a way that there is no possibility that an incident coherent state reaches directly the measured truncated output mode of the whole device, similarly as for the schemes presented in Figs.\ref{f5} and \ref{f6}. Apart from the incident coherent state $|\alpha\rangle$ the interferometer is fed by three other modes being in arbitrary \textit{Fock} states, so that $|\Psi\rangle_{in}=|n_1n_2n_3\alpha\rangle$. At the output ports there are three photon detectors performing photon number measurements. 
The truncation is successful whenever the
number of photons detected equals the number of photons present in the input modes ($n_1,n_2,n_3$) of the whole interferometer. The input state which in fact is a four-mode state, after the measurements is reduced to a single-mode state and has a form of a truncated coherent state in Fock-state representation: $|\Psi\rangle_{out}\cong\sum\limits_{n=0}^{d-1}c_n^{(d)}\gamma_n|n\rangle$ with amplitudes $c_n^{(d)}$ dependent on the beam splitters transmittances and internal phase shifts in the interferometer arms.
The conditions that these parameters have to meet for generation of the  four-, five- and six-dimensional truncated states are derived and presented in \citep*{M05}, as well. 

Apart from the generation of the truncated states up to six-dimensional ones there is also a possibility in that configuration to obtain truncated states with some of $n$-photon states  selectively removed. The Author named that process as a truncation with hole burning  in a Fock space (for the hole burning effect see for instance \citep*{BMB98}). Such states can be written in the $n$-photon states basis as:
\begin{eqnarray}
|\Psi_{without_{(k_1,k_2,...)}}\rangle_{trunc}=N\sum\limits_{\substack{n=0\\(n\neq k_1,k_2,...)}}^{d-1}\gamma_n |n\rangle
\label{eq62}
\end{eqnarray}
where N denotes the normalization constant. Using the eight-port interferometer it can be possible to obtain the states in four dimensional Fock space having the explicit form:
\begin{eqnarray}
|\Psi_{without_{(0)}}\rangle\approx \gamma_1 |1\rangle +\gamma_2 |2\rangle + \gamma_3 |3\rangle\nonumber\\
|\Psi_{without_{(1)}}\rangle\approx \gamma_0 |0\rangle +\gamma_2 |2\rangle + \gamma_3 |3\rangle\\
|\Psi_{without_{(2)}}\rangle\approx \gamma_0 |0\rangle +\gamma_1 |1\rangle + \gamma_3 |3\rangle\nonumber
\label{eq63}
\end{eqnarray}
Each of these states can be formed for various values of the transmittances of beam splitters when the phase shift is arbitrary chosen -- for the specific values see \citep*{M05}. 

\subsection{LQS with atom - cavity field interactions}
There are other methods allowing generation arbitrary Fock states using atom - cavity field interactions.  Among them one should mention a method proposed in \citep*{VAS93}. The Authors consider two-level atoms prepared in a specially chosen superposition of their states interacting  with a cavity field in such a way that only one atom at a time is present in a cavity. This interaction can be described via Jaynes-Cummings model. After the atom has left the cavity its state is measured, and providing that all of the $N$ atoms considered are found to be in their ground states, the truncation is successful. For this case the cavity field state is left in a superposition of $N+1$ number states: $|\Psi^{(N-1)}\rangle=\sum\limits_{n=0}^{N-1}\Psi_n^{(N-1)}|n\rangle$. In such a way a truncated nonclassical state can be created. Although in \citep*{VAS93} an example of the possibility of obtaining a cavity field in one of a truncated phase states is discussed, other states can be also created in their model.

Similar approach is presented in \citep*{GSMC94}. Again various quantum states (among them arbitrary Fock states) can be obtained with the use of the method based on an atom - cavity interactions. Similarly as in \citep*{VAS93}, it is assumed that only one atom at a time interacts with the cavity field where the cavity has Q factor high enough to minimize dissipation during the state preparation. Moreover, the 3-level atoms are involved in a ladder-type configuration where intermediate level is unpopulated during the interaction, so it can be effectively eliminated. The atoms entering the cavity are excited and after the interaction with the cavity field a conditional measurement is performed by the photoionization method. After one atom has passed the cavity and has been found in its ground level the Schr\"odinger cat state in a cavity can be created. It appears that after an interaction of a large number of atoms, the cavity field can remain in a specified number state. Assuming a constant atom-cavity field interaction time, after passage of $N=500$ atoms the field becomes almost a pure number state with a flat phase distribution. The same scheme but with subsequently decreasing interaction times decreases significantly the number of atoms needed for obtaining a number state even for the case when the thermal fluctuations are present.

Next model suitable for generating finite dimensional states, based on the interactions between a coherent field and atoms was proposed in \citep*{KSWW89} and developed in \citep*{MBAM04}. In a high Q-cavity microwave field is prepared in a coherent state whereas the atoms passing through the cavity are prepared in highly excited Rydberg states ($n=51$ \citep*{MBAM04}). During this passage atoms' interaction with the cavity field takes place.  Next, the detection of each of the atoms' states after they have left the cavity is performed. Since a projection onto the field state is done during the measurement one can find the state of the cavity field having a knowledge of detectors measurements. As in \citep*{MBAM04} it was shown that it is possible to create \textit{Fock} states $|2^N\rangle$ where $N$ is the number of atoms left the cavity in an appropriate state. This states generation is characterized by a high accuracy, for instance  for $N=3$ the fidelity of the final state is greater than $98\%$. The Authors named their method as \textit{sculpturing coherent states}.

\subsection{Problem of  imperfections in practical realizations}
The important question arises when dealing with theoretical models of obtaining specific quantum states: whether these schemes are experimentally accessible and what may be the main obstacles in practical realizations. As it was presented in previous sections, linear quantum scissors devices consist mainly of simple optical elements and the main problems which can be found originate from those elements' imperfections. Another problem arises with preparing the initial states (usually single photon ones) for the scissors devices. It is a problem related to the detectors efficiencies and their ability to make measurements of specific photon numbers. In a group of models involving beam-splitters two main groups of LQS schemes where mentioned: with and without interferometric setups. As authors convince, the interferometric setups are more resistant to those imperfections and, hence, the probabilities of detecting a specific truncated state are larger for these schemes.
\subsubsection{Initial state preparation}
The problem of the initial states generation is mainly related to a difficulties in obtaining single-photon states. To overcome those difficulties some authors \citep*{PPB98,BGMB01} suggest using parametric fluorescence and suitable filtering for single-photon state preparation \citep*{BW70}. In \citep*{KKGJ00} making use of non-degenerate parametric down-conversion or when necessary two down-conversion processes were proposed \citep*{CZSJN00,BPMEWZ97,PBWZ98} to produce two-mode $|11\rangle$ state which is used as an initial state for preparing truncation up to two photon states like the one described in (\ref{eq58}).
 
Parametric down-conversion (using a nonlinear crystal) is also used for obtaining an appropriate initial state in the article by \"Ozdemir and coworkers in \citep*{OMKI01}. The theoretical background model for their proposal is that of Pegg-Philips-Barnett \citep*{PPB98} but it is modified so that they focus mainly on its experimental realization and influence of various experimental parameters on efficiency of state truncation.
In addition they discuss the situation of a non-ideal single photon state obtained in down-conversion because of the nonlinear crystal physical properties (size and cone of output radiation effects). They shown that these problems can be solved by spatial and frequency filtering. The authors also take into account in their considerations of an experimental preparation of LQS device a small but finite probability of producing more than one photon in down-conversion process. It is shown in \citep*{OMKI01} that despite the fact of non-perfect initial state preparation there is still a non-vanishing probability that the truncated state will have the desired form.

As a source of higher populated Fock states (like that needed in Villas-B\^{o}as and coworkers proposal \citep*{BGMB01}) one can also make use of already mentioned \textit{Fock} filtering scheme \citep*{AMPS00}.

Other problems which can be met in the practical realization of LQS are related to the preparation of a coherent state, which is used as the second kind of input state. For the schemes discussed here this state is mixed with the entangled state obtained from the incident \textit{Fock} states, and in fact this coherent state expansion in \textit{Fock} states basis is cut by the LQS device. When we apply these LQS procedure the intensity of such input coherent state plays an important role. This intensity should be optimized so that truncation probability is not negligible. This problem is considered in \citep*{OMKI02b}. In order to use the main ideas of the theoretical Pegg-Philips-Barnett scissors scheme the Authors proposed some of its modifications so that it can be experimentally feasible. In particular, an optimalization of coherent state intensity in order to obtain the truncation to zero and one-photon states  was performed. Depending on the detector efficiency the fidelities of obtaining the equally weighted superposition of zero and one-photon states for various coherent state intensities were analyzed. It comes out that even for small detectors efficiencies a proper choice of that intensity guarantees a non-vanishing fidelity for a desired state. From the simulations presented by the Authors one can find that for small intensities of coherent field ($|\alpha|^2\leq 0.36$) the fidelity value does not depend on the detector efficiency, and is always less than 0.84 and further diminishes with decreasing $|\alpha|^2$. So, even while using detectors with maximal efficiency the fidelity for small coherent state intensities will always be lower than 0.84. On the other hand even small detectors efficiencies ($\simeq 0.1$) does not exclude the possibility of detecting the desired truncated state with a high fidelity. For $|\alpha|^2=1.06$ and detector efficiency equal to 0.7 the fidelity of the specified truncated state reaches the value  $\simeq 0.89$. For other combinations of the vacuum and one-photon states, for which  the ratio of the probability amplitudes for obtaining the vacuum and one-photon states differs from $1$, the Authors presented the parameters optimized for obtaining high fidelities. The generalized form of a desired state is:
\begin{equation}
|\Psi\rangle_{cut}=\frac{c_0|0\rangle+c_1|1\rangle}{\sqrt{|c_0|^2+|c_1|^2}}\,\,\, .
\label{eq64}
\end{equation}
and one can choose for the maximal possible fidelity of obtaining arbitrary superposition of zero and one-photon states (specified value of $|c_1/c_0|$ ratio) optimal values of the coherent state intensity and the detector efficiency. The appropriate figures visualizing the best choices are presented in \citep*{OMKI02b}. Their results allow to conclude that for the case of dominant vacuum state in (\ref{eq64}), there is a possibility of achieving fidelities above $0.9$ for relatively small detectors efficiencies. 

In general the ideal model of LQS proposed by Pegg-Philips-Barnett \citep*{PPB98} gives higher fidelities in obtaining truncated states of the form (\ref{eq64}) with a dominant contribution of one-photon state than experimental adoption presented in \citep*{OMKI02b}. However, for the case when the vacuum state is dominant, there are ranges of the values of the parameters that make experimental model more efficient than that proposed in \citep*{PPB98}. For balanced superposition one can also find the parameter ranges for which an experimental approach gives higher fidelities.

Whenever low coherent state intensities are required they can be obtained from the intense coherent states   using optical homodyne tomography as presented for instance in \citep*{SBRF93,SBCR93}.

\subsubsection{Detectors and their imperfections}
The other very important group of problems that can be found when obtaining the truncation with use of LQS are that connected to the non-ideal detectors. These elements are vital in truncation procedure as the appropriate truncated state is obtained providing that the used detectors counted the specified number of photons. Avalanche photodiodes are commonly used as photodetectors and one can include into account the fact that these elements have nonunitary efficiency and there is a nonvanishing probability of non-zero dark-counts. Another problem can be the discrimination between states with different photon numbers. Also the time between the arrival of subsequent photons should be larger than the detector's \textit{dead time} to avoid missing photons. In other cases some photons which achieve the detector sooner than the previous one can be ''noticed '' are simply neglected. In such a situation the truncation cannot be successful. As a method to avoid such situations weak intensity light can be used in an LQS device or small repetition frequencies in pumping the nonlinear crystal (if it is used as a source of single-photon states in a parametric down-conversion process) or in a coherent state preparation.

In \citep*{OMKI01} the authors discuss the influence of detectors imperfections and make a comparison between various photodetectors types  effects on proper truncation. It appears that even though various kinds of detectors inefficiencies can be found, there is still a possibility of obtaining a desired truncated state with high fidelity. The use of conventional photon counters (distinguishing between the presence and absence of photons only) with efficiencies of $\simeq 0.7$ order can produce fidelities above $0.9$ for properly chosen intensity of coherent incident light. For single-photon counters (distinguishing between zero, one and higher than one photons) their efficiencies more remarkably influence the fidelities of proper truncation. They also suffer from a high dark counts rate. However, if the appropriate coherent state intensity is used, the application of combination of these two types of detectors allows for generation of a desired truncated state  -- see \citep*{OMKI01}.

In \citep*{FMC03} the Authors suggest that some problems with detectors efficiencies can be omitted  using a technique in which a light beam is split into many beams and each of them is detected by a standard photodetector device \citep*{HHP04,RHHPH03,ASSBW03,FJPF03}. In that way it is possible to simulate a single-photon resolution. Obtaining such resolutions is a one of the greatest experimental challenges.

\subsubsection{Problems with mode mismatch}
Apart from the other previously mentioned problems in experimental realization of LQS devices there is one which is connected with spatial and temporal matching of the optical fields used in the whole device. 
In order to obtain a successful state truncation and control the form of the state which comes out of the device there should be a perfect matching of all the optical fields both at the beam splitters and at the detectors.  This condition is one of the main assumptions for all theoretical models discussed here. At the beam splitters the fields interfere with each other forming an entangled state. After that a state which is a superposition of the latter with a coherent state is formed. The matching would play an important role in these processes as we want to be sure that the states we observe at the output are the states that we have really wanted.
In a real experiment all of the optical fields incoming to the LQS elements are prepared independently.
This problem was extensively explored in \citep*{OMKI02a} where the influence of temporal mode mismatching on the preparation of a desired input state as well as the interference between these fields and the detectors were considered. The Authors analytically examined these problems in the pulse-mode projection-synthesis approach and found that for  low intensity coherent fields $|\alpha|^2 \ll 1$ both modes mismatch and detector efficiency does not crucially influence the form of the desired state. With increasing value of $|\alpha|^2$ the fidelity for the obtained desired truncated state decreases. For determined values of $|\alpha|^2$ increasing the detector efficiency would lead to the decrease of the fidelity when mode mismatch becomes larger.

\subsection{Quantum scissors and teleportation of the states}
The scheme of the linear quantum scissors is based on the entanglement and the nonlocality phenomenon and, as such, can be applied for the devices used for the teleportation quantum states\citep*{BBC93}.
The truncated state obtained through the projection synthesis is in fact teleported as no light from coherent light mode $|\alpha\rangle$ reaches the mode of the field $|\Psi\rangle_{out}$.  The LQS model of Philips-Pegg-Barnett is therefore sufficient to observe a teleportation effect for the state from the mode of a coherent field to the mode of an output state $|\Psi\rangle_{out}$, and the teleportation effect itself in a system of LQS was a subject of several papers.

\begin{figure}[h!]
\quad\vspace*{2cm}\\
\begin{center}
\includegraphics[scale=0.75]{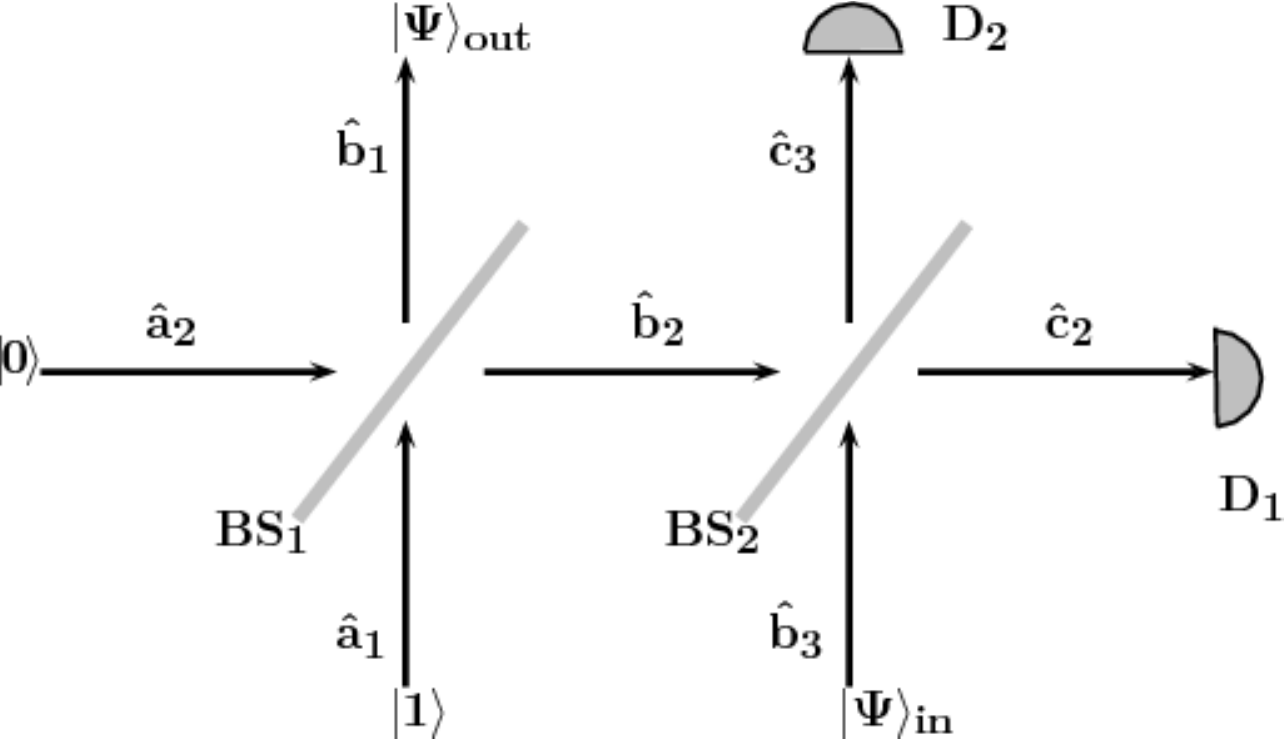}
\end{center}
\caption{Scheme for the linear quantum scissors device involving two beam-splitters $BS_1$ and $BS_2$, and photon detectors $D_1$, $D_2$ \citep{PPB98}. The explanation of operation principle can be found in the text.}
\label{f4}
\end{figure}

\begin{figure}[h!]
\begin{center}
\includegraphics[scale=0.6]{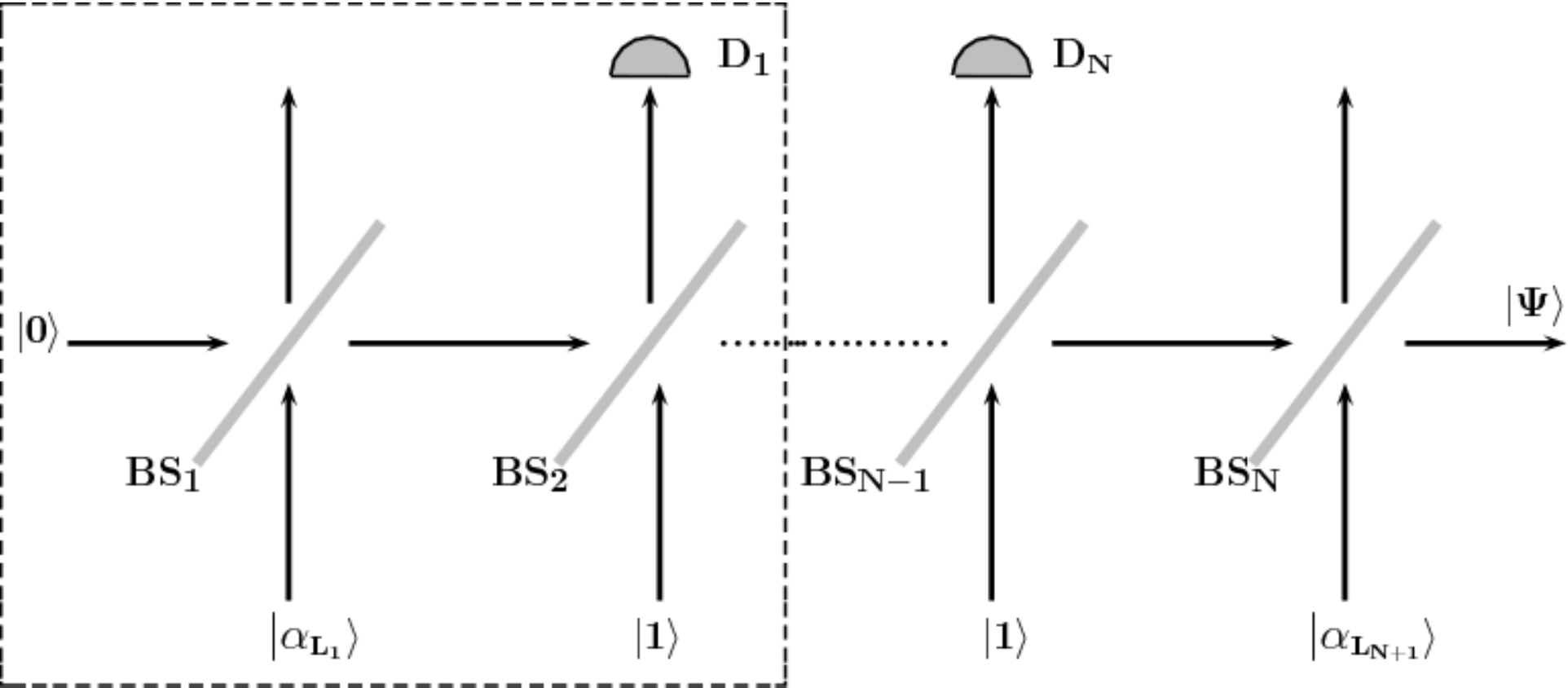}
\end{center}
\caption{Scheme for the linear quantum scissors device composed of a series of beam-splitters; $BS_N$ are beam splitters and $D_N$ are  photon detectors \citep{DCKW99}. The explanation of the operations performed in the model can be found in the text.}
\label{f5}
\end{figure}

\begin{figure}[h!]
\quad\vspace*{2cm}\\
\begin{center}
\includegraphics[scale=0.7]{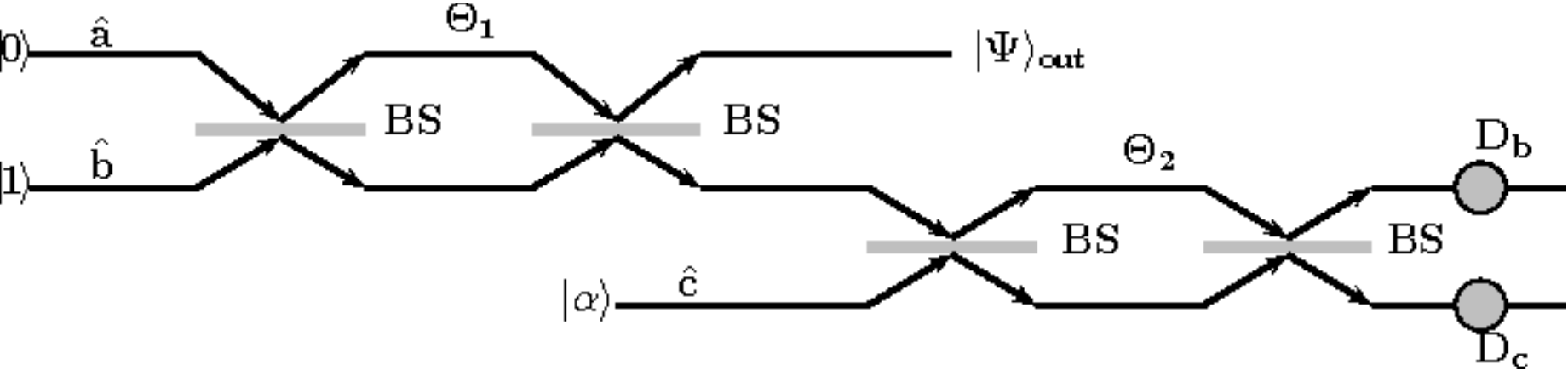}
\end{center}
\caption{Scheme for the linear quantum scissors device composed of two Mach-Zehnder interferometers; $BS$ are beam splitters and $D$ denote photon detectors \citep{P00}. Explanation of the operations performed inside the device can be found in the text. }
\label{f6}
\end{figure}

For example, in \citep*{BAM99} the technique of projection synthesis proposed by Pegg, Philips and Barnett was used to show the possibility of an experimental realization of quantum teleportation of the zero- and one-photon running-wave states by projection synthesis. They explored the effects of losses in beam splitters and the efficiency of detectors on the fidelity of the teleportation. In fact the device they used for teleportation was composed of two Philips-Pegg-Barnett devices like those depicted in Fig.\ref{f4}. The output state of the first device $|\Psi\rangle_{out}$ is introduced to the second one in the place of a coherent field. So, the whole experimental setup was arranged in such a way that the state which is supposed to be teleported by the second device is prepared in the first one by the projection synthesis in a form of a superposition involving vacuum and one-photon states
\begin{equation}
|\Psi\rangle_{out}=N\left(|\gamma_0\rangle_E|0\rangle+|\gamma_1\rangle_E|1\rangle\right)\quad ,
\end{equation}
where the states $|\gamma_{0,1}\rangle_E$ are created as a result of teleportation of a coherent state by the first teleporting device and the influence of the imperfections perturbing the teleportation process.

For the model presented in \citep*{KKGJ00} the Authors showed a possibility of truncating up to two-photon states in an LQS device shown in Fig.\ref{f4}, where some modifications of incoming light fields were introduced. They showed that in their LQS device a teleportation of three-state system is possible. In fact, they discussed a possibility of teleporting of the state $|\Psi\rangle=\gamma_0|0\rangle+\gamma_1|1\rangle+\gamma_2|2\rangle$ which will be the state in mode $\hat{b}_3$ at Fig.\ref{f4}, whereas the form of the state in mode $\hat{b}_1$ is therefore determined by the measurements made by both photodetectors. There is a possibility of successful teleportation, which occurs in the discussed case occurs whenever both detectors count one photon. For the cases when two photons are measured by one of the detectors and none by the second, the state which comes out of the device has different coefficients at each of its \textit{Fock} components.

A direct experimental realization of the Philips-Pegg-Barnett LQS device is done in \citep*{BRL03} by Babichev and co-workers. A maximally entangled state: $|\Psi\rangle=\frac{1}{\sqrt{2}}\left(|1\rangle|0\rangle-|0\rangle|1\rangle\right)$ is prepared at the beam-splitter $BS_1$ and is used for teleporting the state of the coherent light expressed in an infinite dimensional \textit{Fock} state basis. The device, as it was already mentioned, truncates this expansion to the arbitrary superposition of one-mode states, which in fact is simultaneously teleported. The state $|\Psi\rangle_{out}$ that is an output state of the device is then subjected to the homodyne measurement. The accuracy of the teleportation and is fidelity can be determined, and from the experimental data it appears that the fidelity of teleportation achieves high values (up to $99\%$) for small values of the amplitude of the coherent field $\alpha$. For $\alpha=0.5$ this fidelity decreases to $\simeq 80\%$ and for higher amplitude values remarkably decays. However, the fidelity of teleportation is always greater than for the case when the semi-classical limit is achieved.  

\subsection{Multimode states generation -- GHZ and NOON states}
Greenberger-Horne Zeilinger (GHZ) states (see eqn.(\ref{eq51})) are examples of multipartite entangled states with at least three particles involved \citep*{GHS90,GHZ89}. The simplest state of this form is GHZ-state for the three \textit{qubits} case can be written as: $|\Psi\rangle_{GHZ}=\frac{1}{\sqrt{2}}\left(|0\rangle_A|0\rangle_B|0\rangle_C+|1\rangle_A|1\rangle_B|1\rangle_C\right)$. These states are highly non-classical ones and show the multipartite entanglement. The main experimental setup for generating and observing such states relies on the interferometric setup with linear optical elements (beam splitters, polarizing beam splitters, photon detectors) and a nonlinear crystal in which short laser pulse generates pairs of polarization entangled photons. There are four photodetectors, one of them is trigger one, and if they detect photons simultaneously it appears that the other three photons (not counted by a trigger detector) form a polarization GHZ state. The experimental data \citep*{GHZ89} clearly confirm the existence of a three photons polarization entangled state. Moreover, the tests of nonlocality for such states have also been performed \citep*{PBD00}.

There are many papers concerning generation of GHZ states but in this short review we shall concentrate only on a few of them. Especially, we are interested in those that can be qualified as quantum scissors devices. Thus, the proposal based on the atomic system model of the observation of the three-particle entanglement in a cavity was discussed in \citep*{RNO00}. Moreover, the generation of atomic  GHZ states was done in \citep*{ZG01} with two non-resonant vacuum state cavities, and in the study  \citep*{Z01}, where three atoms were passing through one vacuum state cavity. In three atom scheme all the atoms are prepared in the same state $\frac{1}{\sqrt{2}}\left(|g\rangle+|e\rangle\right)$  and therefore there is no energy exchange and the state of the cavities minimizes the decoherence processes. During the collective atom excitations a phase shifts of atomic states appear and creation of an entangled state is possible.

Moreover, quite recently, generation of N-\textit{qubit} GHZ state is reported in \citep*{Y11}. The model presented there consists of a cavity with one three-level and  $N-1$ four-level systems. After an appropriate choice of the atomic transitions which are resonant with an external field and atomic interactions with cavity modes, one can obtain an entangled state of N \textit{qubits}  for some lengths of the interaction times. The model presented there is general enough to be applied not only to the quantum optical systems but also for instance, to the solid-state ones. Obviously, there are only examples of the models that allow for GHZ generation. Other proposals of the generation and manipulation of GHZ states can be found, for example, in the references of the mentioned above papers.

Another quantum multimode state that is finite-dimensional and exhibits entanglement features are the NOON states \citep*{S89}.  They belong to the group of maximally entangled multi-particle ones and for the photon field-states can be expressed as in eqn.(\ref{eq48}). These states are in some sense a generalization of the Bell states, where the one-mode one-photon states $|1\rangle$ is replaced by its $n$-photon counterpart $|N\rangle$. For the quantum optical models the NOON states are generated by photons which are path entangled by optical devices. 

States of these type were discussed for the first time by Sanders \citep*{S89} while considering the decoherence processes in created Schr\"odinger cats. These states were generated by a quantum nonlinear rotator -- quantum two-state system (for example spin system) which performs a quadratic precession around one of its axis. This model is equivalent to the quantum optical nonlinear one and, therefore, should be qualified as \textit{nonlinear} quantum scissors rather than LQS. However, NOON states were obtained in various other schemes, for instance photonic ones discussed in \citep*{KLD02,CK09,CD07,KD07,MLS04} just to name the few of them. Moreover, such states appear in quantum lithography systems, for example see \citep*{BKAB00}.

At this point we should mention one of the latest experimental examples of obtaining ''high-NOON'' states comes from Afek and co-workers \citep*{AAS10}, where the states of the (\ref{eq48}) type with $N=5$  were generated by interfering appropriately in a Mach-Zehnder interferometer classical coherent light with the field obtaining from a spontaneous parametric down-conversion. After the first beam splitter the desired path-entangled states are generated in the both modes. At the output of the second beam splitter two photon detectors measure coincidences. The Authors obtained high values of the overlaps between the experimental results and a theoretical analytical approach. For the states with $N=2$ and $N=3$ the fidelity equal $1$, and for higher NOON states: $N=4$ reaches $0.933$, and for $N=5$  is equal to $0.941$. The experimental setup proposed by the Authors can be used for arbitrary $N$ states providing that proper photon number-resolving photodetectors are used. 

\section{Nonlinear quantum scissors}
Nonlinear quantum scissors (NQS) are a group of optical devices in which nonlinear elements (like nonlinear oscillators) are used \citep{PP00,BDF01}. The effect of action of NQS on the optical state is similar to that previously described for the case of linear quantum scissors (LQS) --- all these devices truncate the optical state which lives in an infinite Hilbert space to the state composed of a few only $n$-photon states completely described in a finite Hilbert space. There have been several proposals of such devices and generally they can be divided into groups differing between themselves by the number of modes of the states obtained. Like in LQS devices the truncation process is inseparably connected with state teleportation and in NQS devices the truncation can be connected with other quantum phenomena. For instance, there is the generation of maximally entangled states or \textit{entanglement death} and \textit{revival} events appearing under special conditions.
One can find examples of NQS which are used for generation of one, two or three modes truncated states, and examples of such models can be found in literature. In this paper we shall subsequently describe the main examples of all of these groups in the following paragraphs.
 
\subsection{NQS devices and one-mode states}

The procedures of preparation single mode \textit{Fock} states are related to preparing quantum scissors devices. Some of those proposals are included in the review work of Dell'Anno, De Siena and Illuminati \citep*{ASI06}. In the group of NQS, which are based on nonlinear optical elements, there are devices which produce single-mode or multi-mode \textit{Fock} states. First we will concentrate on the systems that are able to generate finite-dimensional states in one mode of the EM field. Among them one should mention about the method of ''cutting'' field operators proposed in \citep*{LT94} and developed in \citep*{LDT97}.

As it was mentioned in the first section, the FDCS can be obtained using displacement operator $\hat{D}(\alpha, \alpha^\star)$ acting on the vacuum state, where the boson creation and annihilation operators were defined in a finite-dimensional Hilbert space (see eqns.(\ref{eq19})-(\ref{eq25})), contrary to the TCS, which are defined as an effect of the truncation of the expansion of the Glauber coherent state in $n$-photon states basis (eqns. (\ref{eq26}) and (\ref{eq27})). To generate FDCS we can apply the procedure based on the nonlinear Kerr medium located inside a high-Q cavity and excited by series of ultra-short coherent pulses  \citep*{LT94}. For such a model the unitary operator governing the system's evolution is given by:
\begin{equation}
U=e^{-i(\chi T/2)\hat{n}(\hat{n}-1)}e^{-i(\epsilon\hat{a}^{\dagger}+\epsilon^{\star}\hat{a})}
\label{fock_3}
\end{equation}
where $\chi$ is a nonlinearity constant, $T$ denotes the time between subsequent kicks, $\hat{n}=\hat{a^\dagger\hat{a}}$ is the photon number operator. The parameter $\epsilon$ is in fact an effective strength describing the external field -- nonlinear system interaction. If we assume that the system initially was in the vacuum state and the excitation is weak (\textit{i.e.} $\epsilon\ll \chi$)  it is possible to generate one-photon state with high accuracy, providing that the damping processes are low enough. Moreover, for this case the system's evolution becomes closed within the set of two states: the vacuum $|0\rangle$ and one-photon state $|1\rangle$.

One can modify this model and after the proper changes of the detunings within the system the part of the system's Hamiltonian corresponding to the nonlinear medium evolution becomes
\begin{equation}
\hat{H}=\frac{1}{2}\chi\hat{n}(\hat{n}-2)\quad .
\label{fock_4}
\end{equation}
If we additionally assume that the excitation process is a two-photon one, the system's evolution will be closed within the states $|0\rangle$ and $|2\rangle$. In consequence, for such a situation the two-photon state can be generated. One can generalize this model to that where the nonlinear Hamiltonian will be proportional to $\hat{n}(\hat{n}-z)$, where $z$ is an arbitrary natural number, and the excitation involves $z$-photon process. For this case the $z$-photon state will be generated \citep*{L96}.

A different approach based on construction of a proper Hamiltonian that enables producing an $n$-photon \textit{Fock} state was proposed in \citep*{KH95}, where some transformations from the vacuum state to $n$-photon one and \textit{vice versa} were discussed. These transformation  can be formally written as:
\begin{equation}
\begin{split}
\hat{H}_n|0\rangle&=|n\rangle\\
\hat{H}_n|n\rangle&=|0\rangle \quad .
\end{split}
\label{kilin1}
\end{equation}
In fact, in that paper some subclass of the transformations was discussed. The Authors concentrated on those defined by 
\begin{equation}
e^{it_n\hat{H}_n}|0\rangle=|n\rangle\quad ,
\label{kilin2}
\end{equation}
where $t_n=\pi/2+2\pi k$ and $k$ is an integer number. In consequence, the nonlinear Hamiltonian generating the above transformations was found and it can be written as:
\begin{equation}
\hat{H}_n=\mu\hat{a}^{\dagger}\hat{a}-\mu\frac{\hat{a}^{\dagger}\hat{a}}{n}+\left[\left(1-\frac{\hat{a}^{\dagger}\hat{a}}{n}\right)\frac{\hat{a}^2}{\sqrt{n!}}+H.c.\right]\quad .
\label{fock_5}
\end{equation}
If we assume that the real parameter $\mu=0$ the Hamiltonian represents a physical system that can be practically obtained with use of nonlinear medium. The pumping field of frequency $\Omega$ is within the process described by (\ref{fock_5}) and is turned into the $n$-photon \textit{Fock} state of frequency $\omega=\Omega /n$ in two ways simultaneously: $\Omega\rightarrow n\omega$ and $\Omega+\omega\rightarrow (n+1)\omega$.

Sculpturing a coherent state can also be obtained with use of nonlinear optical elements in a set of Mach-Zehnder interferometers \citep*{MBAM04}. In each interferometer's arms there a cross-Kerr two-mode medium is located, and the interferometer is fed by the vacuum and one photon states. This Kerr medium couples the state of the field in the interferometer with an external field initially prepared in a coherent state. After the field passes the set of interferometers and under the conditions imposed on detections, it ends up in a highly excited \textit{Fock} state provided that the parameters describing Kerr medium were properly chosen. The result of such a system evolution resembles that for the scheme with Rydberg atoms, high-Q cavity and maser field interactions.

Finite dimensional \textit{Fock} states can also be generated in one Mach-Zehnder interferometer with a cross-Kerr medium in one of its arms \citep*{GB06}. Similarly to \citep*{MBAM04} the traveling field is in a coherent state but the interferometer is fed by a vacuum state and another coherent field of larger amplitude. The state inside the interferometer and the external one are coupled via the cross-Kerr medium and after a proper photon detection the filed is found to be in a number state with high accuracy.

\subsection{NQS device for two-mode states}
In the filed of NQS for obtaining two-mode number states one should notice of the first proposals presented in \citep*{LM04,ML06}. The Authors proposed an optical system composed of a nonlinear coupler externally driven by a coherent field. In the core of NQS in this device is a system of two nonlinear oscillators described by the Kerr nonlinearity. There is an interaction between these oscillators and it can be of various character. In \citep*{LM04,ML06} this is a linear interaction, in subsequent papers \citep*{KL06,KL09,KL10a,KL10b,KL10c} this interaction is of nonlinear character whereas in \citep*{KLP11} the interaction is of a parametric type. 

\subsubsection{Nonlinear coupler with linear interaction}
For the linear type of interaction between two oscillators with Kerr-like nonlinearity ({\em nonlinear coupler}) the Hamiltonian governing the systems' dynamics has the following form \citep*{LM04,ML06}:
\begin{eqnarray}
\hat{H}=\hat{H}_0+\hat{H}_{int}=
\underbrace{\frac{\chi_a}{2}(\hat{a}^{\dagger})^2(\hat{a})^2}_{\mbox{$\hat{H}_a$}} + 
\underbrace{\frac{\chi_b}{2}(\hat{b}^{\dagger})^2(\hat{b})^2}_{\mbox{$\hat{H}_b$}} + \underbrace{\epsilon\,\hat{a}^{\dagger}\hat{b}+\epsilon^{\star}\hat{a}\hat{b}^{\dagger}}_{\mbox{$\hat{H}_{int}$}}
+\underbrace{\alpha\,\hat{a}^{\dagger}+\alpha^{\star}\hat{a}}_{\mbox{$\hat{H}_{ext_{a}}$}}+
\underbrace{\beta\,\hat{b}^{\dagger}+\beta^{\star}\hat{b}}_{\mbox{$\hat{H}_{ext_{b}}$}}
\label{nqs_3}
\end{eqnarray}
where $\hat{H}_{int}$ and $\hat{H}_{ext_{a(b)}}$ describe the Hamiltonians for coupler internal and external interaction respectively, and the parameters $\alpha$, $\beta$ are interaction strenghts with external coherent fields which can influence both of the oscillators modes ($\hat{a}$ and $\hat{b}$) or only one of them. Moreover, $\chi_{a(b)}$ are third-order susceptibilities which describe the nonlinearities of the oscillators -- $\hat{H}_{a(b)}$ stand for oscillators Hamiltonians. The general scheme for nonlinear scissors device used in \citep*{LM04,ML06} and also other devices with nonlinear coupling is shown in Fig.\ref{f8}

While considering the no-damping case, the two-mode states which are generated in a process described by the Hamiltonians (\ref{nqs_3}) have the common form:
\begin{equation}
|\Psi(t)\rangle=\sum\limits_{n,m=0}^{\infty}c_{nm}|n\rangle_a|m\rangle_b
\label{nqs_4}
\end{equation}
where the probability amplitudes $c_{nm}$ can be derived from the differential equations:
\begin{eqnarray}
i\frac{d}{dt}c_{nm}&=&\frac{1}{2}\left\lbrace n(n-1)\chi_a+m(m-1)\chi_b\right\rbrace c_{nm}\nonumber\\
&+& \epsilon\,\sqrt{n(m+1)}\;c_{n-1,m+1}+\epsilon^{\star}\sqrt{(n+1)m}\;c_{n+1,m-1}\label{nqs_5}\\
&+&\alpha\,\sqrt{n}\;c_{n-1,m}+\alpha^{\star}\sqrt{n+1}\;c_{n+1,m}+\beta\,\sqrt{m}\;c_{n,m-1}+\beta^{\star}\sqrt{m+1}\;c_{n,m+1}\,\, .\nonumber
\end{eqnarray}

To get the closed set of the above equations (\ref{nqs_4}), and hence, the system's dynamics closed within the finite set of the $n$-photon states, the appropriate conditions should be fulfilled. In general, we should assume that the excitations should be sufficiently weak, \textit{i.e.} the values of the parameters describing these interactions should be considerably smaller as we compare them with the nonlinearity constants. Thanks to those assumption and as a result of the degeneracy of the Hamiltonians $\hat{H}_{a(b)}$ there is a possibility to generate only few two-mode states. In fact in the situation of the NQS device presented in \citep*{LM04,ML06} for the initial state being the vacuum one, the whole dynamics is closed within the system of four two-mode states only. The truncated state has now the simple form:
\begin{equation}
|\Psi(t)\rangle_{trunc}=c_{00}(t)|0\rangle_a|0\rangle_b + c_{01}(t)|0\rangle_a|1\rangle_b + c_{10}(t)|1\rangle_a|0\rangle_b + c_{11}(t)|1\rangle_a|1\rangle_b\, .
\label{nqs_6}
\end{equation}
From the quantum computation theory point of view one can conclude that the two-\textit{qubit} system is produced in evolution of such NQS system. The accuracy of such truncation was proven by the analysis of fidelity between the states (\ref{nqs_4}) and (\ref{nqs_6}). Any deviations from the perfect truncation are of the order of $10^{-4}$ for properly chosen parameters, especially the ratios between the nonlinearities and interaction constants ($\alpha$, $\beta$, $\epsilon$) --- see \citep*{ML06}.

The states which survive after truncation process are the states that form the Bell basis (see eqn.(\ref{eq47})). It was found there that apart from the truncation itself the entanglement is additionally generated. For the NQS pumped in one mode only all of the possible Bell-states can be periodically generated providing that the strength of external coupling is smaller than that corresponding to the internal coupling between two oscillators.  For such situation the states:
\begin{eqnarray}
|B\rangle_1 &=&\frac{1}{\sqrt{2}}\left(|1\rangle_a |1\rangle_b +i|0\rangle_a |0\rangle_b\right)\;\;\; \qquad
|B\rangle_2 =\frac{1}{\sqrt{2}}\left(|0\rangle_a |0\rangle_b +i|1\rangle_a |1\rangle_b\right)\label{nqs_7}\\
|B\rangle_3 &=&\frac{1}{\sqrt{2}}\left(|0\rangle_a |1\rangle_b -i|1\rangle_a |0\rangle_b\right)\;\;\; \qquad
|B\rangle_4 =\frac{1}{\sqrt{2}}\left(|1\rangle_a |0\rangle_b -i|0\rangle_a |1\rangle_b\right)\nonumber
\end{eqnarray} 
are generated. While both strengths are of the same order, generation probability for the states $|B\rangle_1$ and $|B\rangle_2$ becomes close to the unity. For the case when we assume that the system is pumped in the two modes, only the states $|B\rangle_1$ and $|B\rangle_2$ have significant contribution to global entanglement even for $\epsilon > \alpha,\beta$. The degree of entanglement produced by the whole system can be determined by various parameters. For instance, it can be von Neumann entropy, which for the general situation described by the Hamiltonians (\ref{nqs_3}) has the form \citep*{ML06}:
\begin{equation}
E(t)=\epsilon\left(2\left|c_{00}(t)c_{11}(t)-c_{01}(t)c_{10}(t)\right|\right)
\label{nqs_8}
\end{equation}
where $c_{ij}(t)$ are the $|i\rangle_a|j\rangle_b$ states amplitudes, which are obtained from the solutions of the appropriate Schr\"odinger equation. In both situations described ($\beta=0$ and $\beta\neq 0$) the evolution of the entropy (\ref{nqs_8}) clearly points out that the system can be a generator of maximally entangled states.

When assuming more realistic situation of a damped system, it is necessary to apply master equation approach with appropriate damping constants $\gamma_a$ and $\gamma_b$ describing the loses in both coupler modes. For the amplitude damping, the general form of such equation is given by \citep*{GZ00}:
\begin{eqnarray}
\frac{d\hat{\rho}}{dt}&=&-i\left[\hat{H},\hat{\rho}\right]+\frac{\gamma_a}{2}\left(\left[\hat{a}\hat{\rho},\hat{a}^{\dagger}\right]+\left[\hat{a},\hat{\rho}\hat{a}^{\dagger}\right]\right)+\frac{\gamma_b}{2}\left(\left[\hat{b}\hat{\rho},\hat{b}^{\dagger}\right]+\left[\hat{b},\hat{\rho}\hat{b}^{\dagger}\right]\right)\label{nqs_9}\\
&+&\gamma_a\bar{n}_a\left[\left[\hat{a},\hat{\rho}\right],\hat{a}^{\dagger}\right]+\gamma_b\bar{n}_b\left[\left[\hat{b},\hat{\rho}\right],\hat{b}^{\dagger}\right]\nonumber
\end{eqnarray}
or for the case of the phase damping (without loosing the energy in cavities) it becomes:
\begin{eqnarray}
\frac{d\hat{\rho}}{dt}&=&-i\left[\hat{H},\hat{\rho}\right]+\frac{\gamma_a}{2}\left(2\bar{n}_a+1\right)\left[2\hat{a}^{\dagger}\hat{a}\hat{\rho}\hat{a}^{\dagger}\hat{a} -\left(\hat{a}^{\dagger}\hat{a}\right)^2\hat{\rho} - \hat{\rho}\left(\hat{a}^{\dagger}\hat{a}\right)^2\right]\label{nqs_10}\\
&+&\frac{\gamma_b}{2}\left(2\bar{n}_b+1\right)\left[2\hat{b}^{\dagger}\hat{b}\hat{\rho}\hat{b}^{\dagger}\hat{b} -\left(\hat{b}^{\dagger}\hat{b}\right)^2\hat{\rho} - \hat{\rho}\left(\hat{b}^{\dagger}\hat{b}\right)^2\right]\nonumber \quad .
\end{eqnarray}
The effect of the amplitude damping on the entanglement survival can now be determined via concurrence \citep*{W98} or negativity \citep*{P96,HHH96}. For both of the describing here NQS devices the effect of the amplitude damping is much serious problem than dephasing one only, even for different from zero thermal photon numbers. Anyway, the entanglement in both cases decreases quickly with time. For the parameters used in \citep*{ML06} the entanglement produced by NQS subjected to amplitude damping oscillates in time, but its first (and the highest) maximum decreases after less than $2\times 10^{-7}$ [s], if we assume that $\chi_a=\chi_b=10^8$ [rad/s], $\alpha=\chi/20$ and $\epsilon=\alpha/2$. However, if only dephasing effects are present (we have no amplitude damping effects), this time does not change considerably, but the values of the entanglement obtained are much higher and even for larger values of the damping parameters -- the entropy for the first maximum is $\simeq 0.8$ which suggest significant amount of entanglement.

For the NQS scheme with a linear interaction between the Kerr-like oscillators the degree of squeezing in each mode was also studied. In \citep*{SWU07} the Authors estimated squeezing via the quadrature variances for the field in $k$-mode as:
\begin{eqnarray}
\langle \Delta\hat{X}_{k}^{2}\rangle &=& \langle\hat{X}_{k}^{2}\rangle - \langle\hat{X}_{k}\rangle ^{2}\label{nqs_11}\\
\langle \Delta\hat{Y}_{k}^{2}\rangle &=& \langle\hat{Y}_{k}^{2}\rangle - \langle\hat{Y}_{k}\rangle ^{2}\nonumber
\end{eqnarray}
From the considerations given in \citep*{SWU07} it turns out that while the system is not externally driven by a coherent field ($\alpha=\beta=0$) the field generated by NQS device is not squeezed. Taking into considerations classical external pumping, the system produces the squeezed states of light for certain moments of time apart from the entangled states as well.

\subsubsection{Nonlinear coupler with nonlinear interaction}
Another group of NQS devices is composed of systems with nonlinear oscillators that interact with each other in the way that differs from that discussed in the previous section. In particular, we shall assume here that this interaction is of nonlinear character. If such an interaction between the two Kerr-like oscillators is set up, there is also a possibility of obtaining two-mode state truncation. The Hamiltonian for a whole system (\ref{nqs_3}) is therefore modified at the point corresponding to the interaction between two oscillators. Thus, the latter can be written as in \citep*{KL06}:
\begin{equation}
\hat{H}_{int}=\epsilon\,(\hat{a}^{\dagger})^2\hat{b}^2+\epsilon^{\star}\hat{a}^2(\hat{b}^{\dagger})^2 \quad .
\label{nqs_12}
\end{equation}
In considerations presented there only one external excitation by a coherent field was assumed ($\beta=0$). As in previous NQS models, the necessary condition $\chi_{a(b)}\gg\epsilon,\alpha$ for successful truncation have to be fulfilled. Under such an assumption and providing that the initial state for NQS is $|2\rangle_a|0\rangle_b$, the system evolves with high accuracy (confirmed by the fidelity analysis) inside the set of only three states from the whole infinite set of the possible $n$-photon states. Thus, for this case the system's evolution can be described by the following truncated wave-function:
\begin{equation}
|\Psi(t)\rangle_{trunc}=c_{20}(t)|2\rangle_a|0\rangle_b + c_{12}(t)|1\rangle_a|2\rangle_b + c_{02}(t)|0\rangle_a|2\rangle_b \, .
\label{nqs_13}
\end{equation}
Here, one can say that can be treated as a \textit{qutrit}-\textit{qubit} one. It appears that maximally entangled states are generated during the system's evolution:
\begin{eqnarray}
|B\rangle_1&=&\frac{1}{\sqrt{2}}\left(\left|2\right>_a\left|0\right>_b
+i\,\left|0\right>_a\left|2\right>_b\right)\,\,\,, \nonumber \\
|B\rangle_2&=&\frac{1}{\sqrt{2}}\left(\left|2\right>_a\left|0\right>_b
-i\,\left|0\right>_a\left|2\right>_b\right)\,\,\,,\label{nqs_14}\\
|B\rangle_3&=&\frac{1}{\sqrt{2}}\left(\left|2\right>_a\left|0\right>_b
+i\,\left|1\right>_a\left|2\right>_b\right)\nonumber
\end{eqnarray} 
and they are Bell states again. The states $|B\rangle_{1,2}$ are generated giving the $100\%$ entanglement, whereas $|B\rangle_3$ are produced with lower accuracy, however, for this state  the degree of entanglement is higher than $90\%$.

To describe the interactions of this NQS model with an external environment, the master equation approach is used again. As a measure of entanglement degree in the whole \textit{qutrit}-\textit{qubit} system the negativity \citep*{P96,HHH96,VW02} was used. It is defined  as
\begin{equation}
N(\rho)= max\left(0,-2\lambda_{min}\right) \quad ,
\end{equation}
where $\lambda_{min}$ is the smallest of the eigenvalues of the partially transposed system's density matrix.
Examples of the negativity evolution were calculated for various values of the damping constants and are presented in Fig.\ref{f9}. Under no damping condition ($\gamma=0$) the maxima visible in the plot corresponds to the Bell states creation and the second of them (lower maximum) corresponds to the formation of the entangled state $|B\rangle_3$. Between the times in which the system forms one of the Bell-like states (\ref{nqs_14}) negativity does not decay to zero values -- this is because $N(\rho)$ measures the degree of entanglement in the whole system, not only in its \textit{qubit} subspaces. When comparing NQS with a nonlinear internal interaction with NQS containing its linear counterpart described in \citep*{ML06}, the Authors concluded that the former is less sensitive to the damping process and is able to generate states with higher values of entanglement.

There was also considered the influence of the kind of environment interacting with coupler on the entanglement formation in the system. In \citep*{KL09} the Authors considered the evolution of initially entangled state but there \textit{qubit} subspaces were considered. It appears that because of the fact that the whole system has more than two dimensions it is possible to observe complete vanishing and after some time reappearing of the entanglement in the system's two-dimensional subspaces (Fig.\ref{f10}). These effects are known as \textit{sudden deaths} and \textit{revivals of the entanglement}. They were described for the first time for entanglement evolution of two-level atoms (\textit{qubits}) as a result of interactions with external environment \citep*{ZHHH01,YE04,FT06}. The possibility of the observation of such \textit{sudden deaths} and \textit{revivals} in a considered system is the result of the interactions between the oscillators -- the entanglement in the system initially prepared in one of its maximally entangled states (MES) can be transferred from one subspace to another and \textit{vice versa}. Thus, tracking the entanglement in one of these subspaces one can observe such phenomena. Even for interactions with thermal reservoir ($\bar{n}_a\neq0$ and $\bar{n}_b\neq 0$) \textit{deaths} and \textit{revivals} are still present.

What is interesting, for the cases when even the mutual interactions are absent, there is still a possibility for maintaining the entanglement providing that one of the modes is coherently pumped. In particular, when the system is initially prepared in $|B\rangle_1$ state, the coherent external pumping in $\hat{a}$ mode guarantees a continuous energy transfer to(from) the system and in consequence, the entanglement is able to come back to the subspace in which was previously established and from which it was pumped out. This phenomenon appears despite the fact that the damping is present --  see examples in Fig.\ref{f11}. If we assume that the damping is related to the thermal reservoir such behavior is changed in such a way that some dark periods appear in entanglement evolution \citep*{KL10b} -- Fig.\ref{f12}.

At the end, we should mention that the squeezed reservoir has also been considered as an external environment for NQS with nonlinear mutual interactions \citep*{KL10a}. It appears that this kind of environment influences the entanglement decay in such a way that shortening or extension of the time of disentanglement can be observed depending on the values of the parameters describing squeezed environment. The main decisive factor for changing the disentanglement character is squeezed reservoir phase. For a specified mean number of photons in squeezed vacuum one can observe a significant shortening of the disentanglement time for squeezed vacuum phases close to $\pi$.

\subsubsection{Nonlinear coupler with parametric interaction}

Two-mode truncated states can also be produced in a nonlinear system with parametric interactions. The latter plays a double role. It couples two oscillators and excites them as well.  An example of such a device was given in \citep*{KLP11}. The Hamiltonian describing the system described there is given by:
\begin{equation}
\hat{H}=\frac{\chi_a}{2}(\hat{a}^\dagger)^2\hat{a}^2+\frac{\chi_b}{2}(\hat{b}^\dagger)^2\hat{b}^2+
g\hat{a}^\dagger\hat{b}^\dagger+g^{\star}\hat{a}\hat{b} ,
\label{nqs_param1}
\end{equation}
where $g$ is a parameter describing the strength of the parametric process involved in the system. It is obvious that the number of states present in the system's dynamics depends on the value of $g$. If $g$ is sufficiently small (if compare with nonlinearity $\chi$) the system lives in a subspace cut to the three two-mode states $|0\rangle_a|0\rangle_b, |1\rangle_a|1\rangle_b, |2\rangle_a|2\rangle_b$, assuming that the initial system's state is the vacuum one in the both modes. At this point one can say that the system behaves as a \textit{qutrit}-\textit{qutrit} one. However, it should be stressed out that larger values of coupling $g$ result in a larger dimension of the system's space. Similarly as for the previously mentioned NQS devices, the system presented in \citep*{KLP11} also gives the possibility of creation of maximally entangled Bell-like states. The entanglement can be created in various two-\textit{qubit} subspaces and  can be transferred from one subspace to another. When considering interaction with an external environment, it is possible to observe vanishing of entanglement in one of the subspaces (spanned on $|0\rangle_a|0\rangle_b$ and  $|2\rangle_a|2\rangle_b$ states) for some period of time and after that, its return and stay at the certain level. It is related to the probabilities of appearance of appropriate coherences and to changing the value of the ratio between these coherences and state populations.

The same model can also be a source of not only Bell-like states but also so called ''generalized Bell states'' (see eqn.(\ref{eq49})) defined as in \citep*{BBC93}. When the value of interaction parameter $g$ is smaller than the nonlinearity constants $\chi_{a,b}$ the dynamics can be closed within some of the generalized Bell states. For the same parameters as in \citep*{KLP11} one can conclude that the states $|0\rangle_a|0\rangle_b$ $|1\rangle_a|1\rangle_b$ and  $|2\rangle_a|2\rangle_b$  are the most populated. Therefore, three of the states (\ref{eq49}) can be generated in the model of nonlinear oscillators with a parametric interaction having the particular form (\ref{eq50}). In fact, the system is a generator of two-\textit{qutrit} entanglement. The fidelity parameter corresponding to the generation of generalized Bell states for no-damping case achieves its maximal values $\simeq 0.9$ exhibiting its periodic behavior. When we assume that the damping processes are present the fidelity vanishes with time but some amount of entanglement remains within the system (negativity $\simeq 0.05$) even for long times.

\subsection{NQS devices and three-mode states}
There is another group of truncated states that can be obtained with the use of devices comprising nonlinear elements -- three mode truncated states. In \citep*{SWU06} the Authors indicate a possibility of generating $W$ states of the type \citep*{DVC00}:
\begin{equation}
|W\rangle=\frac{1}{\sqrt{3}}\left(|001\rangle +|010\rangle +|100\rangle\right)
\label{nqs_15}
\end{equation}
in nonlinear coupler based device. $W$ states can be generated, for example, the same way as in the Zeilinger at.al. proposal \citep*{ZHG92}, in which an incident light beam is split by a set of partially reflecting mirrors into three beams. But a nonlinear crystal is placed on a path of the primary, already split beam. The beam is ''number filtered'' as it is subjected to up- and down-conversion processes with a filtering located between them. Providing that one photon has been detected, the field can be found in a $W$ entangled state. Another method \citep*{GZ02} relies on the cavity quantum electrodynamics processes. In this model three two-level atoms that are appropriately prepared interact with a field in a cavity. The cavity is initially in a vacuum state, the same as in Jaynes-Cummings model, or, in a second approach, a single excited two-level atom sequentially interacts with three cavities prepared initially to be in vacuum states. The resultant states (those of atoms or cavities) are the desired $W$ states. Parametric down-conversion processes with linear optical elements can be also used to prepare polarization-entangled $W$ states \citep*{YTKI02}. For this model the parametric process produces four photons which are after that spatially divided via beam splitters into four paths. After a measurement done by the photodetector, the state of the photons is found to be entangled $W$ state.

The NQS device can also be used for obtaining $W$ states.
For that purpose NQS is composed of three Kerr-like oscillators which are mutually connected via the interaction Hamiltonians  \citep*{SWU06}:
\begin{equation}
\hat{H}_{int}=\epsilon\hat{a}^{\dagger}\hat{b}+\epsilon^{\star}\hat{a}\hat{b}^{\dagger}+
\epsilon\hat{a}^{\dagger}\hat{c}+\epsilon^{\star}\hat{a}\hat{c}^{\dagger}+
\epsilon\hat{b}^{\dagger}\hat{c}+\epsilon^{\star}\hat{b}\hat{c}^{\dagger}
\label{nqs_16}
\end{equation}
The truncated wave-function that was obtained is composed of the eight three-mode states: $|000\rangle$, $|001\rangle$, $|010\rangle$, $|011\rangle$, $|100\rangle$, $|101\rangle$, $|110\rangle$ and $|111\rangle$. While the initial state is $|001\rangle$ it appears that with high accuracy the $W$ state is generated. Its explicit form is given in \citep*{SWU06}:
\begin{equation}
|\Psi(t)\rangle=\frac{1}{3}\left[2\exp\left(i\epsilon t\right)+\exp\left(-2i\epsilon t\right)\right]|001\rangle+ \frac{1}{3}\left[-\exp\left(i\epsilon t\right)+\exp\left(-2i\epsilon t\right)\right]\left(|010\rangle+|100\rangle\right)
\label{nqs_17}
\end{equation}
and is generated periodically at times 
\begin{equation}
t_k=\frac{\pi}{3\epsilon}\left[\left(k-\frac{1+(-1)^k}{2}\right)+\frac{1}{3}\left(-1\right)^k\right]\quad ,\quad  (k=1,2,\cdots)\quad .
\end{equation}

The presence of an additional coupling with external fields leads to the possibility of generation of other truncated three-mode states. As for the two-mode state truncation method described in \citep*{ML06} there is also a possibility  of observing squeezed light produced in NQS device for truncating three-mode states \citep*{SWU07}. For the latter, when the external driving field is present, squeezing can be observed. For such a case the squeezing related to the quadrature $\hat{X}$ (\ref{nqs_11}) is present in the modes which are not externally driven, while the pumped modes exhibit squeezing for the quadrature $\hat{Y}$.

NQS can be a source of the NOON states, as well. As it was noted in the previous section concerning LQS devices, the model of quantum rotator proposed by  Sanders \citep*{S89} can be used for the  Schr\"odinger cats and the NOON states generation. The model discussed there is equivalent to that described by the following nonlinear Hamiltonian \citep*{S89}
\begin{equation}
\hat{H}_{int}/\hbar=\omega_a\hat{a}^\dagger\hat{a}+\omega_b\hat{b}^\dagger\hat{b}+2\chi_{ab}\hat{a}^{\dagger}\hat{a}\hat{b}^{\dagger}\hat{b}+\frac{1}{2}\chi_a\left(\hat{a}^{\dagger}\hat{a}\right)^2+\frac{1}{2}\chi_b\left(\hat{b}^{\dagger}\hat{b}\right)^2
\label{noon_2}
\end{equation}
describing quantum optical system with the nonlinear coefficients $\chi_i$ ($i=a,b,ab$) that are components of the tensor $\chi^{(3)}$ for the nonlinear medium -- for simplicity the relation $\chi\equiv\chi_a=\chi_b=\chi_{ab}$ is applied. For this model two modes of the field first pass the nonlinar media and interact with each other during the time $t_1$. Next, they pass through the two separable Kerr media in such a way that these modes do not interact with each other \citep*{SM89} undergoing self-phase shifts during the time $t_2$. The whole interaction scheme is equivalent to the action of the unitary evolution operator
\begin{equation}
\begin{split}
\hat{U}\,=&\,\exp \big\{
-i(\omega_a\hat{a}^\dagger\hat{a}+\omega_b\hat{b}^\dagger\hat{b})\,(t_1+t_2)\\
&-\frac{i}{2}\chi\,\big[(\hat{a}^\dagger\hat{a})^2+(\hat{b}^\dagger\hat{b})^2\big]\,(t_1+t_2)
-2i\,\chi\,\hat{a}^\dagger\hat{a}\hat{b}^\dagger\hat{b\,}t_1
\big\}\quad 
\end{split}
\end{equation}
and it can be used as the NOON states generator.

\subsection{Quantum scissors -- some experimental aspects}
All the NQS devices based on elements with Kerr-like nonlinearities have some experimental restrictions. The most important one is that related to the fact that the excitations should be much weaker as we compare them with the value of the nonlinearities involved in the models. Therefore, for successful (two- or three-mode) state truncation the main assumption $\chi_{a(b)}\gg\epsilon,\alpha,\beta, g$ has to be fulfilled. In practice, such assumption is often equivalent to the \textit{giant Kerr nonlinearities} application. In fact, this obstacle can be overcome in various ways. 

One can notify that for the described here models, the nonlinearity constant $\chi$ appears together with the nonlinear evolution time $T$, so the above condition should be modified in such a way that $\chi$ will be replaced by $\chi T$. Therefore, by the application of the sufficiently long period of time $T$ one can fulfill the requirements necessary for the desired states generation. It can be done, for instance, by the nonlinear fibers of the appropriate length application. For example, a nonlinear fiber that was $300$m long and embedded into a Sagnac interferometer allowed for up to $10^7$ photon pairs per $1$mW of the optical pumping generation \citep*{LVS05}. Similar photon-pair generation rates have also been reported from the experiments in which $2$m-long microstructured fibers were applied\citep*{FAW05}.

The other way is to generate so called \textit{giant Kerr nonlinearities}. Such nonlinearitites can be achieved in atomic systems where quantum interferences are present along with the electromagnetically induced transparency (EIT) effect. For instance, Schmidt and Immamo\u{g}lu \citep*{SI96} shown that for the three- and four-levels schemes driven by the properly prepared external laser field such nonlinearities can be generated. Such models were also discussed in various aspects in \citep*{ISWD97,GWG98,ISWD98}, and their realization in \citep*{KZ03,HHDB99}. Moreover, these effects can be also observed in other than optical systems. For instance, quite recently in \citep*{RTM09} the method of high nonlinearities generation was described for a system of multi-level Cooper pair box molecule interacting with superconducting resonator. 

It should be stressed out that in fact the quantum scissors effect is observable in various models exhibiting so-called \textit{photon blokade} effect. In general we can say that such effect appears when the irradiated by EM field system exhibits photon-photon interaction in such a way that the resonant absorption of one photon ''blocks'' the absorption of the second one. In fact, the latter becomes detuned from the resonance. This effect is an analogue of the \textit{Coulomb blockade} observed in sufficiently small metallic and semiconductor devices. The relations between the \textit{giant Kerr nonlinearities} generation, EIT effect and \textit{photon blockade} was discussed in \citep*{DRT98}. In fact, the \textit{photon blockade} can be observed for the single atom -- cavity models \citep*{BBM05}, nano-cavities ones \citep*{FMV10,LMG10}, Bose-Einstein condensates \citep*{OHP01} or even $XY$ spin systems \citep*{ASB07}.

Another group of models that can be a potential candidate for an NQS practical realization are those comprising cavities with moving mirrors. At this point one should mention one of the early papers concerning these problems \citep*{BJK97}, where such a system was applied for the Schr\"odinger cats and the entangled states generation. It was shown there that the cavity filled with the EM field only and confined by vibrating mirror, can be described by a Hamiltonian with nonlinear terms. Finally, the so called \textit{atomic optics} systems can also be used as NQS. As it was shown in \citep*{WFV97,WV97,WV98,WVK99, MCWV99} trapped and atoms and/or ions can be described by the Hamiltonians that are counterparts of those describing usual quantum nonlinear systems.

\section{Summary}
Finite-dimensional states are commonly applied and discussed  in various quantum-optical, quantum information theory or more generally -- quantum engineering models. One can find various recipes for the \textit{Fock}, finite dimensional coherent, squeezed or entangled states generation in optical systems. \textit{Quantum scissors} are the devices which enable obtaining such states. Depending on the physical properties of the ''device'' used for the state truncation, quantum scissors can be divided into two groups: one of them (LQS) are based on linear optical elements whereas the other (NQS) are constructed with the application of the nonlinear ones. The presented here ideas and models are not only theoretical ones. As a result of the current progress in both: experimental techniques and physical theory, described in this paper models become more realistic and even some of them are accessible for the experiment now.

As it was shown here, quantum scissors can be a source of the variety of quantum states that can be \textit{finite-dimensional} or \textit{truncated} states. These states exhibit various very interesting properties that can be a subject of a broad range of investigation in the field of quantum optics. Nevertheless, one should remember that the problems of quantum scissors can be applied to other than optical models. Moreover, for the case of the multi-mode states, the generated states can exhibit very interesting properties related to the quantum entanglement. This fact indicates that the subject related to the quantum scissors is deeply related to the both: practical and theoretical aspects of the quantum information theory, and to the basic questions of the contemporary quantum theory.


\newpage

\begin{figure}[h!]
\begin{center}
\includegraphics[scale=0.5]{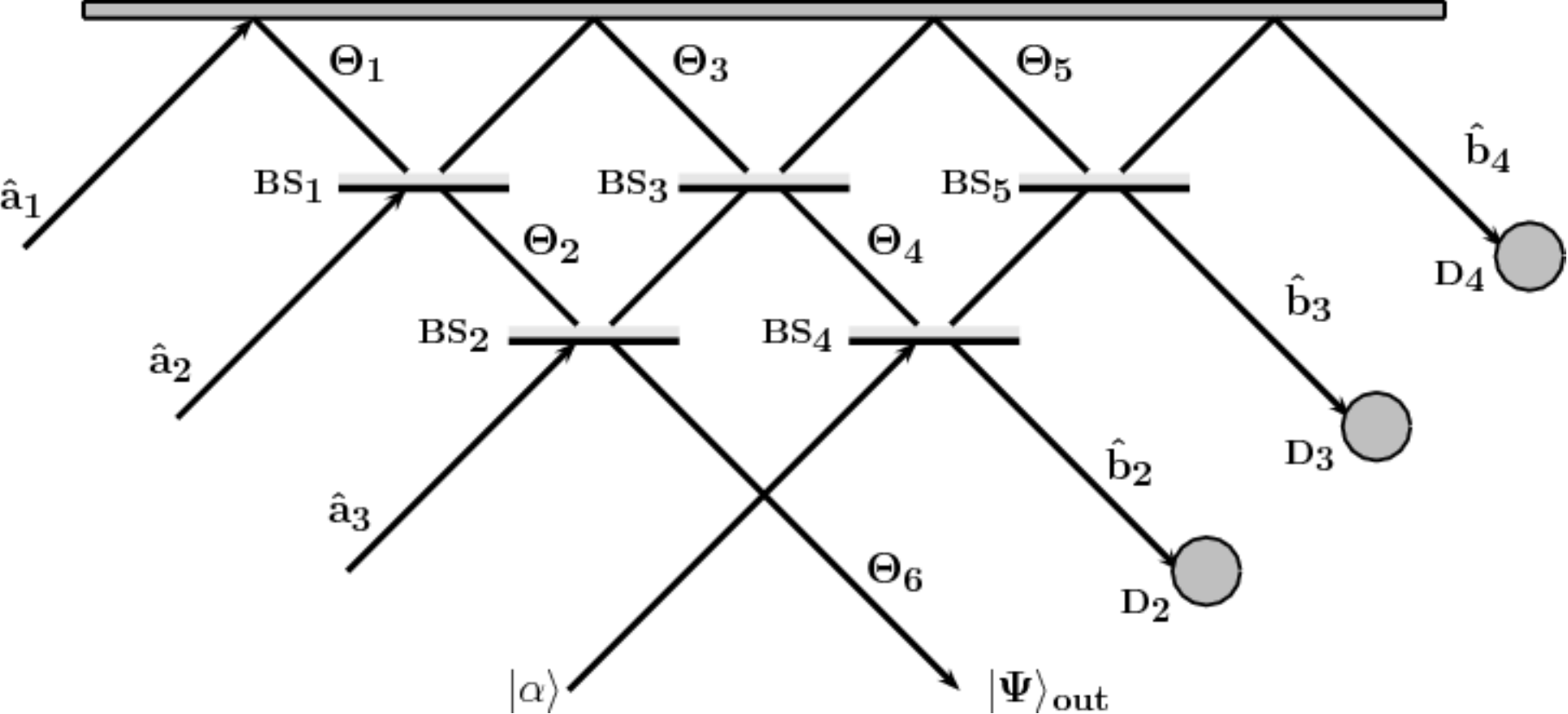}
\end{center}
\caption{Scheme for the linear quantum scissors device composed of the eight-port interferometer; $BS$ are beam splitters and $D$ are photon detectors, $\Theta$ are the phase shifts in interferometer arms \citep{M05}. Explanation of the operations performed inside the device can be found in the text.}
\label{f7}
\end{figure}

\begin{figure}[h!]
\begin{center}
\includegraphics[scale=0.6]{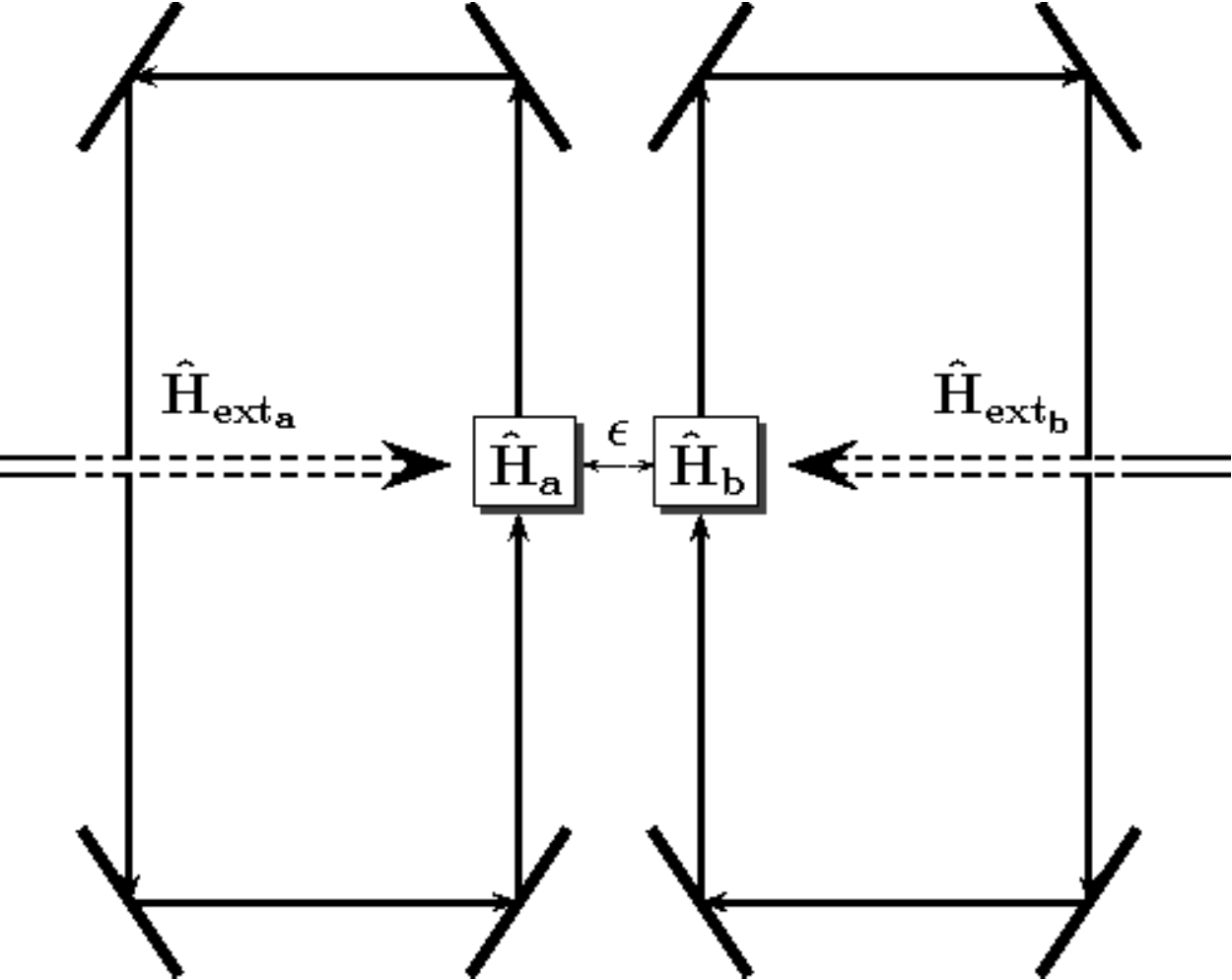}
\end{center}
\caption{General scheme for the two-mode non-linear quantum scissors device.}
\label{f8}
\end{figure}

\begin{figure}[h!]
\begin{center}
\begin{center}
\includegraphics[scale=0.5]{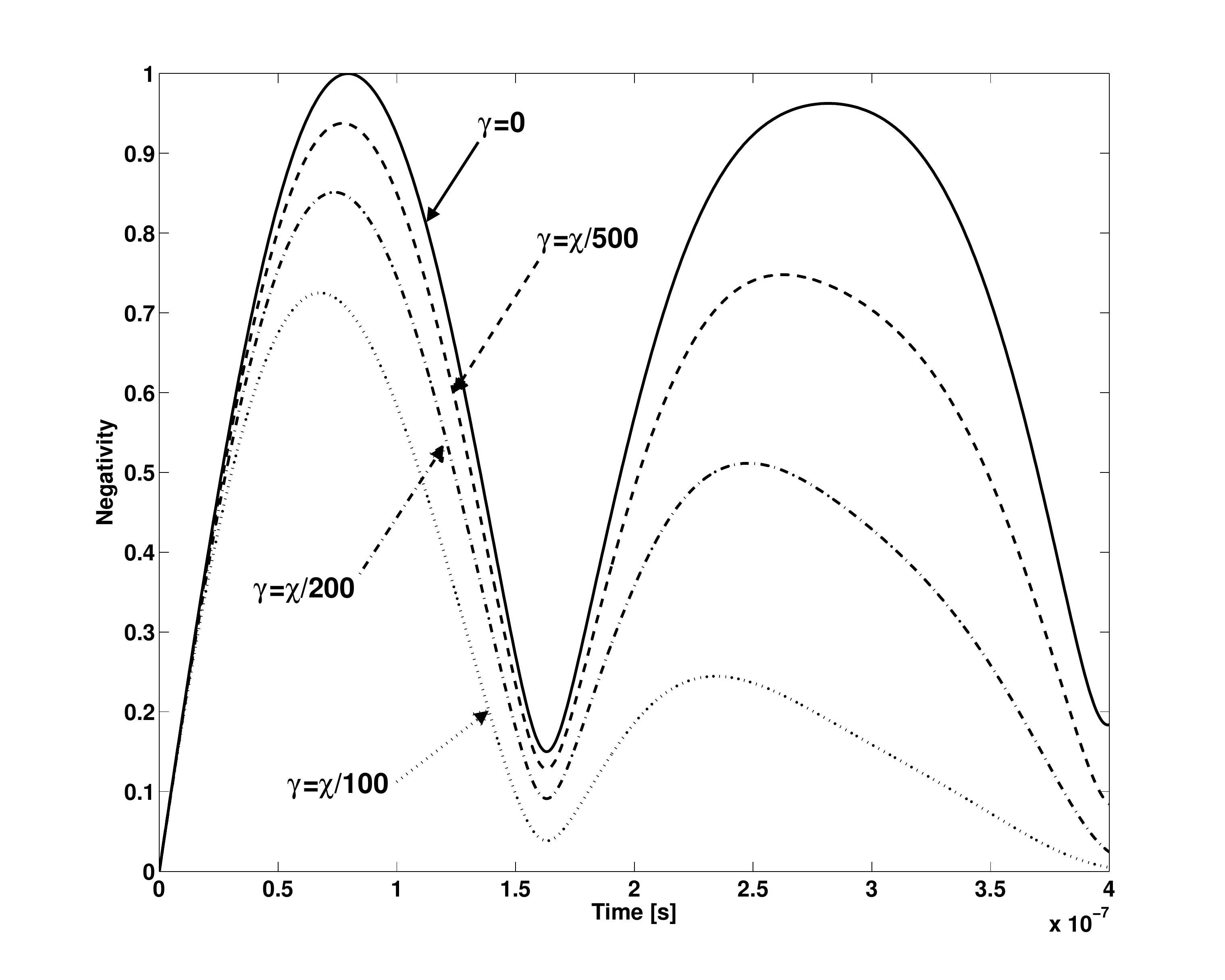}
\end{center}
\end{center}
\caption{Time-dependance of the negativity for the NQS device with a nonlinear interaction between the oscillators for various values of damping constants $\gamma_a=\gamma_b=\gamma$. $\chi_a=\chi_b=\chi=10^8 [rad/s]$, $\alpha=\chi/20$, $\epsilon=\alpha/2$.}
\label{f9}
\end{figure}

\begin{figure}[h!]
\begin{center}
\includegraphics[scale=0.5]{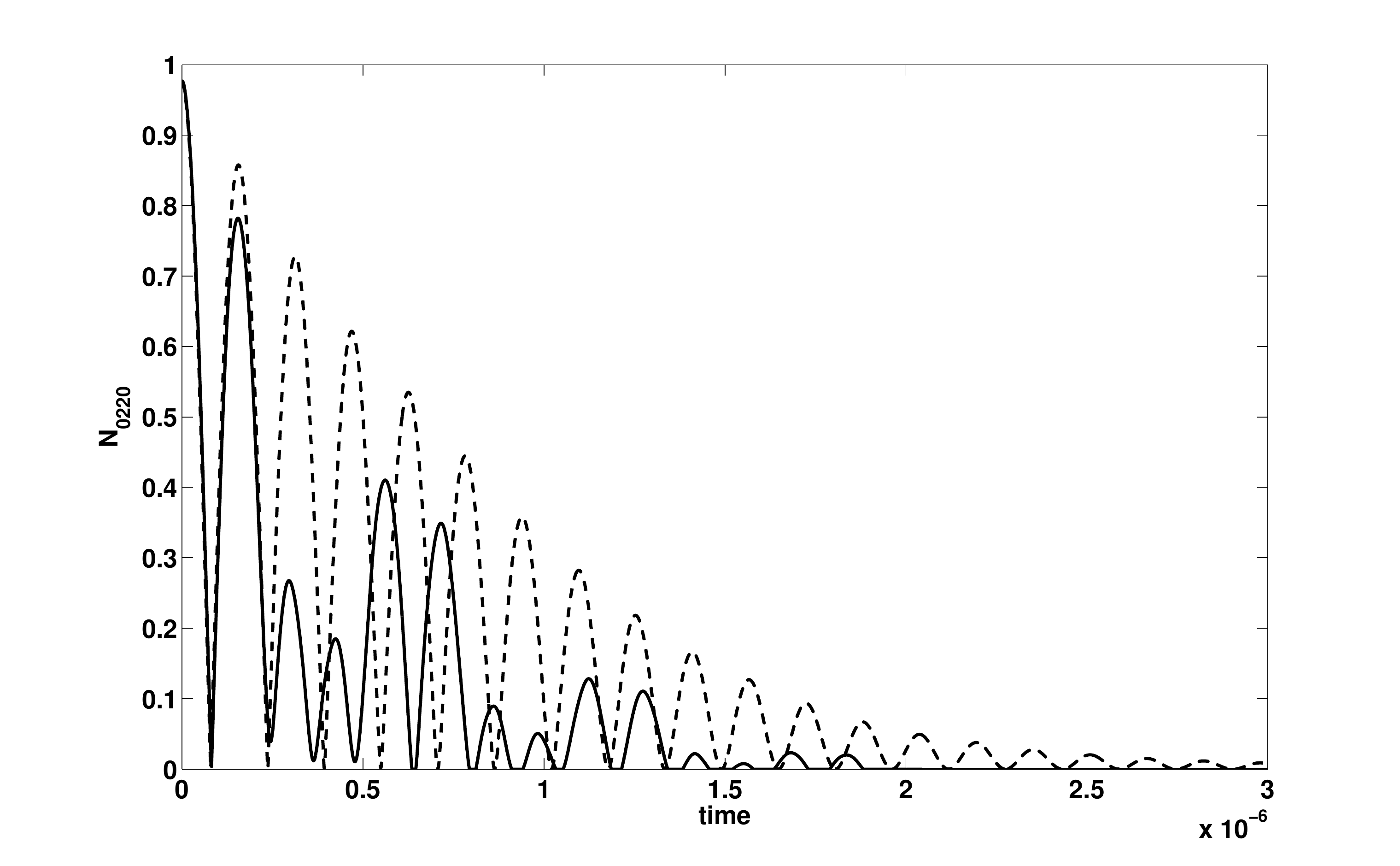}
\end{center}
\caption{Time-dependance of the negativity for the NQS device with a nonlinear interaction between the oscillators for various values of thermal photon numbers. $\chi_a=\chi_b=\chi=10^8 [rad/s]$, $\epsilon=\chi/20$, $\alpha=\epsilon$, $\gamma_a=\gamma_b=\gamma=\chi/500$. Solid line: $\bar{n}_a=0.2$, $\bar{n}_b=0$; dashed line: $\bar{n}_a=\bar{n}_b=0$. Negativity is obtained for the subspace $|0\rangle|0\rangle$, $|0\rangle|2\rangle$, $|2\rangle|0\rangle$ and $|2\rangle|2\rangle$. Initial state is $|B\rangle_1$ state.}
\label{f10}
\end{figure}

\begin{figure}[h!]
\begin{center}
\includegraphics[scale=0.5]{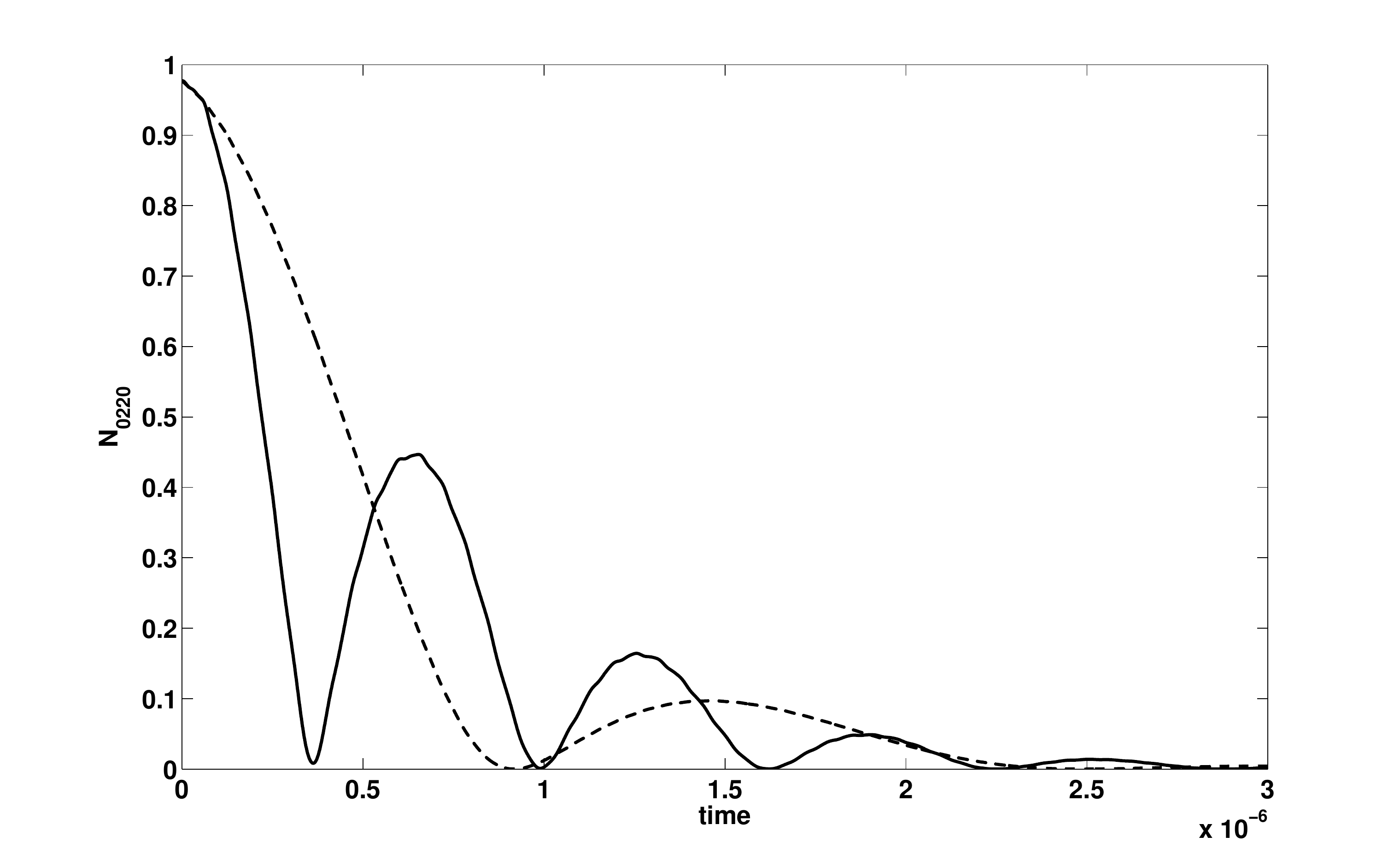}
\end{center}
\caption{The negativity as a function of time for the NQS device with a nonlinear interaction between the oscillators for the zero-temperature bath reservoir and various values of the $\alpha$: solid line -- $\alpha=\chi/20$, dashed line -- $\alpha=\chi/50$. The parameters $\chi_a=\chi_b=\chi=10^8 [rad/s]$, $\epsilon=0$, $\gamma_a=\gamma_b=\gamma=\chi/500$,  $\bar{n}_a=\bar{n}_b=0$.  Negativity is obtained for the subspace $|0\rangle|0\rangle$, $|0\rangle|2\rangle$, $|2\rangle|0\rangle$ and $|2\rangle|2\rangle$. Initial state is $|B\rangle_1$ state.}\label{f11}
\end{figure}

\begin{figure}[h!]
\begin{center}
\includegraphics[scale=0.5]{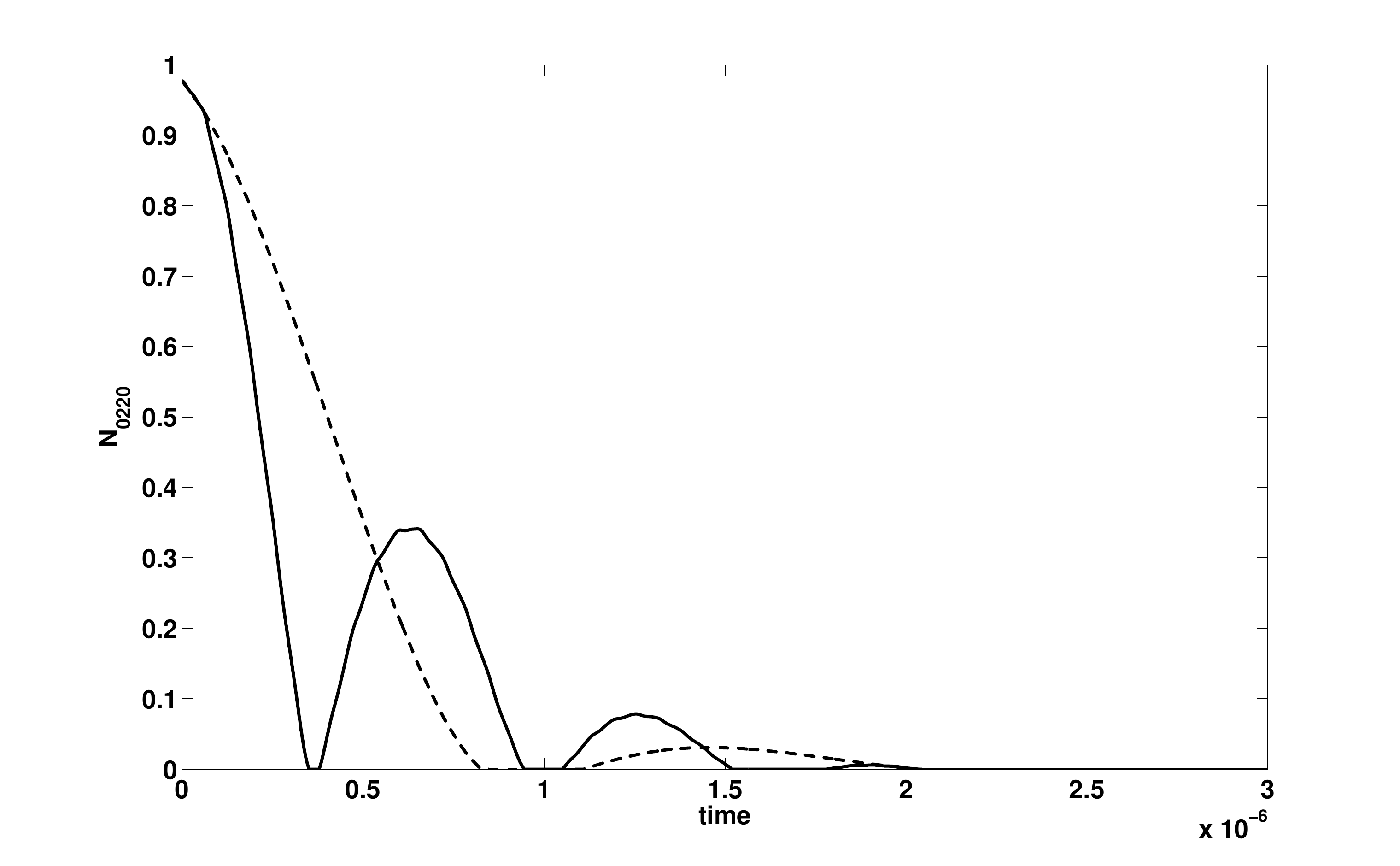}
\end{center}
\caption{Negativity as a function of time for the NQS device with a nonlinear interaction between the oscillators for $\bar{n}_a=0.2$ $\bar{n}_b=0$ and various values of $\alpha$: solid line -- $\alpha=\chi/20$, dashed line - $\alpha=\chi/50$. The remaining parameters are: $\chi_a=\chi_b=\chi=10^8 [rad/s]$, $\epsilon=0$, $\gamma_a=\gamma_b=\gamma=\chi/500$. Negativity is obtained for the subspace $|0\rangle|0\rangle$, $|0\rangle|2\rangle$, $|2\rangle|0\rangle$ and $|2\rangle|2\rangle$. Initial state is $|B\rangle_1$ state.}
\label{f12}
\end{figure}

\end{document}